\documentclass[showpacs,twocolumn,aps,prd]{revtex4-2}
\usepackage{graphicx}
\usepackage{dcolumn}
\usepackage{bm}
\usepackage{amsmath}
\usepackage{amssymb}
\usepackage[usenames,dvipsnames]{xcolor}
\usepackage{color}
\usepackage[colorlinks,plainpages=false]{hyperref}

\begin{document}

\title{%
Matching-invariant running quark masses in 
Quantum Chromodynamics
      }
\author{H.~M.~Chen,$^1$ L.~M.~Liu,$^1$ J.~T.~Wang,$^1$ M.~Waqas,$^1$ G.~X.~Peng,$^{1,2,3}$
       }
\affiliation{
{$^1$School of Physics $\&$ College of Nuclear Science and Technology,}\linebreak
{
University of Chinese Academy of Sciences, Beijing 100049, China}\\
{$^2$Synergetic Innovation Center for Quantum Effects and Applications
} \linebreak
{Hunan Normal University, Changsha 410081, China}\\
$^3$\mbox{Theoretical Physics Center for Science Facilities,
    Institute of High Energy Physics,  Beijing 100049, China}
            }

\begin{abstract}
The renormalization group equations of the QCD running coupling
and quark masses are solved in a mathematically strict way.
General relations between the standard expansion coefficients and
the beta and gamma functions have been accordingly deduced.
It is thus found that previous truncated expressions either loses
an infinite number of known logarithmic terms,
which can be given in a closed form,  or includes a large number of
incorrect terms.
Consequently, a new expansion given by Peng with the fastest convergence speed
is  adopted. Given the matching-invariant beta and gamma functions
to four-loop level, the scale parameters for the coupling and masses
are calculated from the new expressions and the latest experimental
average of $\alpha_s(M_{\mathrm{Z}})$,
for the order from 1 to 4 and the number of flavors from 3 to 6.
In addition, the conventional quark mass is not continuous at thresholds. In this paper, we derive matching-invariant quark masses which are continuous everywhere. They are expanded as an obvious function of the logarithmic $\Lambda$ scaled energy.
The expansion coefficients are related to the original $\gamma$\ and $\beta$ functions, with concretization to four loop level. The results show that the new expressions for the quark masses converge indeed much faster.
\end{abstract}


\maketitle

\section{Introduction}

As is well known, the coupling and quark masses in Quantum
Chronmodynamics (QCD) \cite{Schramm2002}
are running, i.e., they depend on
the renormalization point. To have full thermodynamic consistency
in practical applications \cite{pgx05EPL72,Fraga05PRD71},
it is necessary to include the renormalization-group (RG) dependence
of the coupling and masses.

In nuclear and particle physics, it is usually convenient to have an explicit
expression of the coupling as a function of the renormalization point $u$.
The standard approach is to expand it to a series of
$1/\ln(u^2/\Lambda^2)$, where $\Lambda$\ is a QCD scale parameter
and customarily called $\Lambda_{\mathrm{QCD}}$.
Another approach to calculating the coupling at a given $u$ is to truncate
the beta series in the RG equation, and then solve the corresponding
truncated differential equation numerically.
The relation between the standard expansion coefficients
and the beta function can be found to order 3 in Ref.\ \cite{Eidelman2004PLB592.1},
and to order 4 in Ref.\ \cite{Chetyrkin1997PRL79}.

In the standard minimum subtraction scheme \cite{tHooft1973NPB61.455}
or its modified version  \cite{BardeenPRD18} ($\overline{\mbox{MS}}$),
the QCD coupling is, in principle, not continuous at the quark thresholds,
which is surely inconvenient for applications. However, a new
matching-invariant coupling can be given by absorbing
loop-effects into the $\overline{\mbox{MS}}$ definition \cite{pgx06plb634}.
For this the beta function should be linearly recombined. The general
relation between the new and original beta coefficients will be
discussed, with concrete expressions to 4-loop level.
The same thing is also true for quark running masses. But, instead of
changing the beta function, one should modify the gamma function.

The running quark masses are conventionally expressed as a series of the
running coupling. Because the relation between the coupling
and the renormalization point is not exactly known, the conventional
expression causes indirect imprecision.

The paper is organized as follows. In Sec.\ \ref{sol-coupliing},
the RG equation for the coupling is solved in a mathematically strict way,
and the general relation between the standard expansion coefficients
and the beta function is accordingly obtained.
In Sec.\ \ref{alf-compar}, the new expression given by Peng is compared with the conventional approaches to show that the new expression converges much faster \cite{pgx06plb634}.
Section \ref{secqkmass} and \ref{mat-alf} works with the matching condition
of the coupling. In addition,
a matching-invariant coupling is derived in Sec.\ \ref{mat-invar-alf}.
The relation between the corresponding new and original beta and gamma
functions is deduced.
Section \ref{exp-qm} is dedicated to the quark running masses.
The RG equation for the masses is solved mathematically
with the corresponding gamma function for the
matching-invariant masses given to order 4.
The quark masses are expressed first as a function of the coupling, then
expanded to a direct series of the logarithmic renormalization point.
Comparison with the conventional expansion and the truncation in the
RG equation show that the new expression has the fastest convergence speed.
Finally a summary is presented in Sec.\ \ref{secsum}.

\section{solution of the RG equation for the coupling}
\label{sol-coupliing}

The renormalization-group equation for the QCD running coupling
$ \alpha=\alpha_s/\pi=g^2/(4\pi^2) $ is
\begin{equation} \label{RGa}
\frac{\mathrm{d}\alpha}{\mathrm{d}\ln u^2} =-\sum_{i=0}^{\infty}
   \beta_i\alpha^{i+2}
\equiv \beta(\alpha),
\end{equation}
where the beta function was calculated, in $\overline{\mbox{MS}}$,
to one-loop level in Ref.\ \cite{Gross1973PRL30.1343},
to two-loop in Ref.\ \cite{Caswell1974PRL33.244},
to three-loop in Ref.\ \cite{Tarasov1980PLB93.429} and
to four-loop in Ref.\ \cite{Ritbergen1997PLB400p379}.
The beta coefficients $\beta_i$ can be expressed as polynomials of
the number of flavors $N_{\mathrm{f}}$, i.e.,
\begin{equation} \label{originalbeta}
\left[
\begin{array}{c}
\beta_0\\
\beta_1\\
\beta_2\\
\beta_3
\end{array}
\right]
=\left[
  \begin{array}{cccc}
  11/4  &  -1/6    &    0     &   0 \\
  51/8  & -19/24   &    0     &   0 \\
\frac{2857}{128} &-\frac{5033}{1152} & \frac{325}{3456} &   0 \\
 \beta_{3,0} &   \beta_{3,1} &  \beta_{3,2}  & \frac{1093}{186624} \\
  \end{array}
          \right]
\left[
 \begin{array}{c}
  1\\
  N_{\mathrm{f}}\\
  N_{\mathrm{f}}^2\\
  N_{\mathrm{f}}^3
\end{array}
\right]
\end{equation}
with
$
\beta_{3,0}=(891/64)\zeta_3+149753/1536
      \approx 114.2303287,\
\beta_{3,1}=-(1627/1728)\zeta_3-1078361/41472
      \approx -27.13394382,\
\beta_{3,2}=(809/2592)\zeta_3+50065/41472
      \approx 1.582379064,\
$ where $\zeta$\ is the Riemann zeta function, and $\zeta_2=\pi^2/6$,
$\zeta_3\approx 1.202056903$, $\zeta_4=\pi^4/90$, $\zeta_5\approx 1.036927755$.
In this paper, all color factors are given for $N_{\mathrm{c}}=3$.
Comparing the beta expressions here with those
in Ref.\ \cite{Ritbergen1997PLB400p379},
one would find a difference by a factor of $4^{i+1}$
due to the form of Eq.\ (\ref{RGa}).

Previously, to solve Eq.\ (\ref{RGa}), one first replaces the infinity
there with a finite order, say 3, and then expand the coupling to a series
and seek the corresponding expansion coefficients. Here the problem
is treated in a different way \cite{pgx06plb634}.

Summarizing the derivations in Ref.\ \cite{pgx06plb634},
we give the general solution of the RG equation 
Eq.\ (\ref{RGa}) of the strong coupling as
\begin{equation} \label{ageneral}
\alpha
=
 \frac{L}{\beta_0}
 \sum_{i=0}^{\infty}
 \sum_{j=0}^i
  f_{i,j}
 \left(\frac{\beta_1L}{\beta_0^2}\right)^i
 \ln^jL,
\end{equation}
where $L=1/\ln(u^2/\Lambda^2)$.
The expansion coefficients $f_{i,j}$
are
\begin{equation} \label{fijrecur}
f_{i,j}
 =
   \frac{1}{(i-j)!}
   \sum_{l=0}^{i-j} f_{i-j,l}
   \sum_{s=0}^j
   \frac{(i-s)!}{(j-s)!}
   \bigsqcup_1^l \frac{1}{s}.
\end{equation}
For the meaning of $\bigsqcup_1^l\frac{1}{s}$,
see Eqs.\ (\ref{lna8})-(\ref{lna11}). Because $\bigsqcup_1^l\frac{1}{s}=0$
when $s<l$, the lower limit of the summation index $s$ can be restricted to
$s=l$, and the upper limit of the summation index $l$ can be constrained to
$\min(i-j,j)$. For explicit results to order 6, please see in Ref.\ \cite{pgx06plb634}.

\section{Finite-form expressions of the running coupling}
\label{alf-compar}

Generally, $f_{i,j}$ involves $\beta_k$'s until $k\le i$.
In reality, we can only know the beta function to a finite
order, say, $N$. Therefore, in perturbation theory
we can only have an approximation of the exact solution.
The question is now how to truncate the series so that it
has the fastest convergent speed. In this section,
we first discuss the two conventional approaches, and then
develop another expansion which converges much faster.

\subsection{Standard truncation}

The conventional approach corresponds to simply replacing the $\infty$\ in
Eq.\ (\ref{ageneral}) with $N-1$, giving
\begin{equation} \label{alfappr01}
\alpha_N^{\mathrm{cv}}(u)
=
 \frac{1}{\beta_0\ln\frac{u^2}{\Lambda^2}}
 \sum_{i=0}^{N-1}
 \left(\frac{\beta_1}{\beta_0^2\ln\frac{u^2}{\Lambda^2}}\right)^i
 \sum_{j=0}^i
  (-1)^j
  f_{i,j}
  \ln^j\ln\frac{u^2}{\Lambda^2},
\end{equation}
where the subscript cv indicates that it is the conventional standard expression.
One can write it to arbitrary order according to the coefficients
in the previous section.
To order 5, for example, we explicitly write
\begin{eqnarray} 
\frac{\beta_0}{L}\alpha_5^{\mathrm{cv}}
&=&
  1
  +\frac{\beta_1 L}{\beta_0^2}\ln L
  +\frac{\beta_1^2 L^2}{\beta_0^4}
   \left(
    U_2
    +\frac{\beta_0\beta_2}{\beta_1^2}
   \right)
   \nonumber\\
&&
   +\frac{\beta_1^3 L^3}{\beta_0^6}
   \left[
     U_3
    +\left(3\frac{\beta_0\beta_2}{\beta_1^2}-2\right)\ln L
    +\frac{\beta_0^2\beta_3}{2\beta_1^3}
   \right]
  \nonumber\\
&&
+\frac{\beta_1^4 L^4}{\beta_0^8}
 \left[
  U_4
 +\left(
    6\frac{\beta_0\beta_2}{\beta_1^2}
   -\frac{3}{2}
  \right)\ln^2 L
   \right.
   \nonumber\\
&& \phantom{+\frac{\beta_1^4 L^4}{\beta_0^8}[}
 +\left(
    2\frac{\beta_0^2\beta_3}{\beta_1^3}
   +3\frac{\beta_0\beta_2}{\beta_1^2}
   -4
  \right)\ln L
 \nonumber\\
&&\left.
  -3\frac{\beta_0\beta_2}{\beta_1^2}
  -\frac{\beta_0^2\beta_3}{6\beta_1^3}
  +\frac{\beta_0^2}{3\beta_1^4} \left(5\beta_2^2+\beta_0\beta_4\right)
 \right].
\end{eqnarray}
where
$U_2=\ln^2 L+\ln L -1$,
$U_3=\ln^3 L+(5/2)\ln^2 L-1/2$,
$U_4=\ln^4 L+(13/3)\ln^3 L+7/6$.
Because $\beta_4$ is not available presently, we can exactly
calculate $\alpha_N^{\mathrm{cv}}(u)$, at most, to $N=4$:
\begin{eqnarray}
\alpha_4^{\mathrm{cv}}(u)
&=&\frac{1}{\beta_0\ln(u^2/\Lambda^2)}
  -\frac{\beta_1\ln\ln(u^2/\Lambda^2)}
         {\beta_0^3\ln^2(u^2/\Lambda^2)}
 \nonumber\\
&&
 \hspace{-1.5cm}
+\frac{\beta_1^2}{\beta_0^5\ln^3(u^2/\Lambda^2)}
 \left[
  \left(\ln\ln\frac{u^2}{\Lambda^2}
                                 -\frac{1}{2}\right)^2
  +\frac{\beta_0\beta_2}{\beta_1^2}
  -\frac{5}{4}
 \right]
  \nonumber\\
&&
 \hspace{-1.5cm}
-\frac{\beta_1^3}{\beta_0^7\ln^4(u^2/\Lambda^2)}
 \left[
 \left(\ln\ln\frac{u^2}{\Lambda^2}-\frac{5}{6}\right)^3
  -\frac{\beta_0^2\beta_3}{2\beta_1^3}
  +\frac{233}{216}
  \right.\nonumber\\
&&
 \hspace{-1.5cm}
 \phantom{-\frac{\beta_1^3}{\beta_0^6\ln^3(u^2/\Lambda^2)}[}
  +\left(3\frac{\beta_0\beta_2}{\beta_1^2}-\frac{49}{12}\right)
   \ln\ln\frac{u^2}{\Lambda^2}
 \Bigg].
\end{eqnarray}
Here the terms are grouped into four parts:
two in the first line, one in the second line,
and another one in the last two lines.
The first one is $\alpha_1^{\mathrm{cv}}$,
the first two give $\alpha_2^{\mathrm{cv}}$,
and the first three are nothing but $\alpha_3^{\mathrm{cv}}$.

\subsection{Truncation in the RG equation}

Another way to obtain an approximation for the coupling
is to truncate the series in Eq.\ (\ref{RGa}) to a definite
order, say $N$, i.e.\
\begin{equation} \label{rgafin}
\frac{\mathrm{d}\alpha}{\mathrm{d}\ln u^2}
=-\sum_{i=0}^{N-1} \beta_i\alpha^{i+2},
\end{equation}
and then solve the corresponding differential equation
exactly.
In this case, we still have the exact series solution in Eq.\ (\ref{ageneral}),
but all the $\beta_{i\ge N}$ should be set to zero.
In the following, we discuss the compact form of the exact
solution of Eq.\ (\ref{rgafin}).

If $N=1$, the solution is easy to get:
\begin{equation} \label{alfaprox1}
\alpha_1^{\mathrm{rt}}
=\frac{1}{\beta_0\ln(u^2/\Lambda^2)},
\end{equation}
which is the same as Eq.\ (\ref{alfappr01}) at $N=1$.

If $N\ge 2$, we define
\begin{equation}
W_N(\alpha)
\equiv
 \beta_0^2
\int_0^{\alpha}
\left[
 \frac{1}{\beta_0 x^2}
 -\frac{\beta_1}{\beta_0^2 x}
 -\frac{1}{\sum_{i=0}^{N-1}\beta_i x^{i+2}}
 \right]
 \mbox{d} x,
\end{equation}
or, equivalently,
\begin{equation}
W_N(\alpha)
\equiv
\int_0^{\alpha}
\frac{
 \sum\limits_{i=0}^{N-2}(\beta_0\beta_{i+2}-\beta_1\beta_{i+1})\alpha^i
 -\beta_0\beta_N\alpha^{N-2}
     }
     {
  \sum_{j=0}^{N-1}\beta_j\alpha^j
     }
\mbox{d}x,
\end{equation}
then Eq.\ (\ref{rgafin}) can be integrated out to give
\begin{equation} \label{intRGa}
\ln\frac{u^2}{\Lambda^2}
= C'
 +\frac{1}{\beta_0\alpha}
 +\frac{\beta_1}{\beta_0^2}\ln\alpha
 +\frac{1}{\beta_0^2}W_N(\alpha).
\end{equation}
To be consistent with the series solution, the constant of
integration should be $C'=\frac{\beta_1}{\beta_0^2} (\ln \beta_0+C)$.
From Eq.\ (\ref{intRGa}), $u$ can be easily
expressed as an explicit function of $\alpha$, i.e.,
\begin{equation} \label{solRGa}
u^2(\alpha)
=\Lambda^2
\exp\left[
          \Theta_N(\alpha)
           +\frac{\beta_1}{\beta_0^2}C
    \right].
\end{equation}
With the inverse function of $\Theta_N$, one then has
an expression for the renormalization scale dependence of
the coupling:
\begin{equation} \label{aexp}
\alpha_N^{\mathrm{rt}}=\Theta_N^{-1}
      \left[
       \ln\frac{u^2}{\Lambda^2}
        -\frac{\beta_1}{\beta_0^2}C
      \right].
\end{equation}
 In Eqs.\ (\ref{solRGa}) and (\ref{aexp}), the function $\Theta_n(\alpha)$
is defined as
\begin{equation} \label{Fnw}
\Theta_N(\alpha)
\equiv
  \frac{1}{\beta_0\alpha}
 +\frac{\beta_1}{\beta_0^2}\ln(\beta_0\alpha)
 +\frac{1}{\beta_0^2}W_N(\alpha).
\end{equation}
The functions $W_N(\alpha)$ are
\begin{equation} \label{alfaprox2}
W_2(\alpha)
=-\beta_1\ln\left(1+\frac{\beta_1}{\beta_0}\alpha\right),
\end{equation}
\begin{eqnarray} \label{alfaprox3}
W_3(\alpha)
&=&
  \frac{2\beta_0\beta_2-\beta_1^2}{\sqrt{4\beta_0\beta_2-\beta_1^2}}
  \arctan\frac{\beta_1+2\beta_2\alpha}{\sqrt{4\beta_0\beta_2-\beta_1^2}}
\nonumber\\
&&
 -\frac{2\beta_0\beta_2-\beta_1^2}{\sqrt{4\beta_0\beta_2-\beta_1^2}}
  \arctan\frac{\beta_1}{\sqrt{4\beta_0\beta_2-\beta_1^2}}
\nonumber\\
&&
 -\frac{\beta_1}{2}
 \ln\sum_{i=0}^2\frac{\beta_i}{\beta_0}\alpha^i,
\end{eqnarray}
and
\begin{equation} \label{alfaprox4}
W_4(\alpha)
=\sum_{i=0,\pm 1}
  \eta_i\ln\left(1+\frac{\alpha}{\alpha_i}\right),
\end{equation}
where $\alpha_i\ (i={0,\pm 1})$ are the three complex roots of
the polynomial equation
\begin{equation} \label{eqa0}
\beta_0-\beta_1 \alpha +\beta_2 \alpha^2 -\beta_3 \alpha^3=0,
\end{equation}
and $\eta_i$ is given by
\begin{equation}
\eta_i
=\frac{\beta_0\beta_2-\beta_1^2
       -(\beta_0\beta_3-\beta_1\beta_2)\alpha_i
       -\beta_1\beta_3\alpha_i^2}
      {\beta_1-2\beta_2\alpha_i+3\beta_3\alpha_i^2}.
\end{equation}

Please note, for the matching-invariant beta coefficients,
which will be given in Eq.\ (\ref{primedbeta}),
Eq.\ (\ref{eqa0}) has one real root and two conjugate complex roots.
The real root is
\begin{equation}
\alpha_0
=\left(t_1+\sqrt{t_1^2+t_2^3}\right)^{1/3}
 -\frac{t_2}{\left(t_1+\sqrt{t_1^2+t_2^3}\right)^{1/3}}
 +\frac{\beta_2}{3\beta_3}.
\end{equation}
with
$
t_1=(1/2)\beta_0/\beta_3
    -(1/6)\beta_1\beta_2/\beta_3^2
    +(1/27)\beta_2^3/\beta_3^3,
$
and
$
t_2=(1/3)\beta_1/\beta_3-(1/9)\beta_2^2/\beta_3^2.
$

The expressions in Eqs.\ (\ref{alfaprox1}), (\ref{alfaprox2}),
and (\ref{alfaprox3}) are all real numbers. The weak point
of Eq.\ (\ref{alfaprox4}) involves calculations in the
field of complex numbers. But the result is a real number
for $W_4(\alpha)$. One can also give an expression to
$W_4(\alpha)$ which concerns only real numbers.
But the result is a little more complicated:
\begin{eqnarray} \label{Wexp4}
W_4(\alpha)
&=&
 \iota_4
 \left[
  \arctan\frac{\alpha-\iota_1}{\sqrt{\iota_0-\iota_1^2}}
  +\arctan\frac{\iota_1}{\sqrt{\iota_0-\iota_1^2}}
 \right]
\nonumber\\
&&
 +\ln\frac{\left(
            1-2\iota_1\alpha/\iota_0+\alpha^2/\iota_0
           \right)^{\iota_3}}
        {\left(1+\alpha/\alpha_0\right)^{\iota_2}},
\end{eqnarray}
where
\begin{subequations}
\begin{eqnarray}
\iota_0
&=&
 \frac{\beta_0}{\beta_3\alpha_0},
\ \
\iota_1
=\frac{\beta_0-\beta_1\alpha_0}{2\beta_3\alpha_0^2}
=\frac{1}{2}\left( \alpha_0-\frac{\beta_2}{\beta_3} \right), \\
\iota_2
&=&
 \frac{\beta_0^2}{\alpha_0
                  (3\beta_0-2\beta_1\alpha_0+\beta_2\alpha_0^2)}, \
\iota_3
=
 \frac{1}{2}(\iota_2-\beta_1),\\
\iota_4
&=&
  \frac{
  \iota_0\left(
           {\iota_2}/{\alpha_0}+\beta_2-{\beta_1^2}/{\beta_0}
          \right)
   +\iota_1(\iota_2-\beta_1)
        }
       { \sqrt{\iota_0-\iota_1^2} }.
\end{eqnarray}
\end{subequations}

To extend Eqs.\ (\ref{aexp}) and (\ref{Fnw}) to $N=1$, one can let
\begin{equation}
W_1=-\beta_1\ln(\beta_0\alpha).  
\end{equation}

When we have $W_N(\alpha)$, the coupling is then obtained
by numerically solving the one-variable equation
\begin{equation}
\frac{1}{\beta_0\alpha}+\frac{\beta_1}{\beta_0^2}\ln(\beta_0\alpha)
+\frac{1}{\beta_0^2}W_N(\alpha)
=\ln\frac{u^2}{\Lambda^2}-\frac{\beta_1}{\beta_0^2}C.
\end{equation}

For $N=2$, $\Theta_N^{-1}(\alpha)$ can also be expressed explicitly by using
the physical branch of the Lambert function \cite{Gardi1998JHEP9807}.
For higher $N$, however,  it becomes difficult.

\subsection{A new expression}

As an application of the general relation between $f_{i,j}$ and
the beta coefficients, let's study another expansion which converges
much faster. For this we rewrite the
general series solution in Eq.\ (\ref{ageneral}) as
\begin{equation} \label{alfexpn}
\alpha=
 \frac{\beta_0}{\beta_1}
 \sum_{i=0}^{\infty}
 {L^*}^{i+1}
 \sum_{j=0}^i
 f_{i,j} \ln^jL
\equiv \sum_{i=0}^{\infty} J_i.
\end{equation}
Representing the terms in this expansion with the corresponding
coefficients $f_{i,j}$, all the terms can be
arranged in a matrix as
\begin{equation} \label{amat}
[\alpha]
=\left[
  \begin{array}{cccccccc}
   1       & 0       & 0       & 0       & 0       & 0 &  \\
   0       & 1       & 0       & 0       & 0       & 0 &  \\
   f_{2,0} & 1       & 1       & 0       & 0       & 0 &  \\
   f_{3,0} & f_{3,1} & 5/2     & 1       & 0       & 0 &\vdots  \\
   f_{4,0} & f_{4,1} & f_{4,2} & 13/3    & 1       & 0 &  \\
   f_{5,0} & f_{5,1} & f_{5,2} & f_{5,3} & 77/12   & 1 &  \\
   \phantom{xxxx} &\phantom{xxxx} & \cdots
   & \phantom{xxxx} &\phantom{xxxx} & \phantom{xxx} &
  \end{array}
 \right].
\end{equation}
The standard expansion corresponds to summing the terms row by row
from the rho 0 to $N-1$.
Then, obviously, an infinite number of terms,
such as $f_{j,j}$ ($j\ge N$) on the diagonal and $f_{j+1,j}$ ($j\ge N$)
on the next to diagonal etc, are lost, though these terms are all known
as in Eq.\ (\ref{fijrecur}).
Generally, the terms $f_{j+k,j}$ on the $k$th next to diagonal involves
only $\beta_{i\le k}$. But all the terms $f_{j+k,j}$
with $k<N$ and $j\ge N$ are known and lost, though no such terms
are zero even if one sets all $\beta_{i>1}$ to zero.


To include the contribution from the terms just mentioned,
Peng consider to sum over diagonals,
which can be achieved by taking $i=j+k$ in Eq.\ (\ref{alfexpn}), then, the following formula is obtained through mathematical derivation.
\begin{equation} \label{alfgen2}
\alpha=
\frac{\chi_0}{\beta_0}
\sum_{k=0}^{\infty}
\sum_{l=0}^k
  f_{k,l} {\chi}^k\ln^l \kappa.
\end{equation}
where
\begin{eqnarray}
\kappa &\equiv&
\frac{1}{1-L^*\ln L}
=\left[
  1+\frac{\beta_1\ln\ln(u^2/\Lambda^2)}{\beta_0^2\ln(u^2/\Lambda^2)}
 \right]^{-1},
\label{kappadef}\\
\chi_0
&\equiv& L\kappa
=\frac{\beta_0^2}{\beta_0^2\ln(u^2/\Lambda^2)+\beta_1\ln\ln(u^2/\Lambda^2)},\\
\chi &\equiv&
\label{chidef}
\frac{\beta_1}{\beta_0^2} \chi_0
=\frac{\beta_1}
  {
  \beta_0^2\ln(u^2/\Lambda^2)+\beta_1\ln\ln(u^2/\Lambda^2)
  }.
\end{eqnarray}

Then, the new expression for the QCD running coupling can rewrite as
\begin{equation} \label{alfgen3}
\alpha_N^{\mathrm{fs}}=
\frac{\beta_0}{\beta_1}
\sum_{k=0}^{N-1}
 \left(
  \frac{\beta_1}{\beta_0^2}
  \kappa L
 \right)^{k+1}
\sum_{l=0}^k
 f_{k,l} \ln^l \kappa.
\end{equation}
To leading order, it is simply
\begin{equation}  \label{alfgen31}
\alpha_1^{\mathrm{fs}}
=
 \frac{\kappa}{\beta_0\ln(u^2/\Lambda^2)}
=
\frac{\beta_0}
     {\beta_0^2\ln(u^2/\Lambda^2)+\beta_1\ln\ln(u^2/\Lambda^2)}.
\end{equation}
To the next to leading order, it becomes
\begin{eqnarray}
\alpha_2^{\mathrm{fs}}
&=&
\frac{\beta_0}
     {\beta_0^2\ln(u^2/\Lambda^2)+\beta_1\ln\ln(u^2/\Lambda^2)}
\nonumber\\
&&
-\frac{\beta_0\beta_1
   \ln\left[
       1+\frac{\beta_1\ln\ln(u^2/\Lambda^2)}{\beta_0^2\ln(u^2/\Lambda^2)}
      \right]}
     {\left[
       \beta_0^2\ln(u^2/\Lambda^2)+\beta_1\ln\ln(u^2/\Lambda^2)
      \right]^2}
\nonumber\\
&=&
 \frac{\kappa}{\beta_0\ln(u^2/\Lambda^2)}
 +\frac{\beta_1\kappa^2\ln\kappa}{\beta_0^3\ln^2(u^2/\Lambda^2)}.
\end{eqnarray}
To order 3, it is
\begin{eqnarray}
\alpha_3^{\mathrm{fs}}
&=&
 \frac{\kappa}{\beta_0\ln(u^2/\Lambda^2)}
 +\frac{\beta_1\kappa^2\ln\kappa}{\beta_0^3\ln^2(u^2/\Lambda^2)}
\nonumber\\
&&
 +\frac{
  \beta_1^2\kappa^3
  \left[
   \left(\ln\kappa+\frac{1}{2}\right)^2
   +\frac{\beta_0\beta_2}{\beta_1^2}-\frac{5}{4}
  \right]
      }
      {
  \beta_0^5\ln^3(u^2/\Lambda^2)
      }
      .
\end{eqnarray}
Presently we can give it to order 4:
\begin{eqnarray}
\alpha_4^{\mathrm{fs}}
&=&
 \frac{\beta_0}{\beta_1}
   \Bigg[
   \chi
   +{\chi}^2\ln \kappa
   +{\chi}^3\left(\ln^2\kappa+\ln\kappa+\frac{\beta_0\beta_2}{\beta_1^2}-1\right)
\nonumber\\
&&
  {\chi}^4
  \left(
   \ln^3\kappa +\frac{5}{2}\ln^2\kappa +f_{3,1}\ln \kappa +f_{3,0}
  \right)
  \Bigg],
\end{eqnarray}
where $f_{3,0}$ and $f_{3,1}$ are given in Ref.\ \cite{pgx06plb634},
$\kappa$\ is in Eq.\ (\ref{kappadef}), and $\chi$\ in Eq.\ (\ref{chidef}).
For convenience, they can be listed here:
$f_{3,0}=(\beta_0^2\beta_3/\beta_1^3-1)/2$,
$f_{3,1}=3\beta_0\beta_2/\beta_1^2-2$,
$
\kappa=\left[
  1+(\beta_1/\beta_0^2)\ln\ln(u^2/\Lambda^2)/\ln(u^2/\Lambda^2)
 \right]^{-1},
$\
and
$
\chi
=\beta_1/
 \left[
  \beta_0^2\ln(u^2/\Lambda^2)+\beta_1\ln\ln(u^2/\Lambda^2)
 \right].
$

\subsection{Comparison of the three expressions}

Now we have three expressions for the coupling, i.e.,
Eqs.\ (\ref{alfgen3}), (\ref{aexp}), and (\ref{alfappr01}).
These expressions can be generally represented by
\begin{equation} \label{agenthree}
\alpha
=\alpha_N^{(i)}(u,N_\mathrm{f},\Lambda,C),
\end{equation}
where $i=$ cv, rt, and fs represent, respectively, the expressions
in Eqs.\ (\ref{alfappr01}), (\ref{aexp}), and (\ref{alfgen3}).
Therefore, the coupling depends not merely on the renormalization
point $u$, but also on $i, N, N_\mathrm{f}$, i.e.,
for a theoretical value of the coupling, one must know it was calculated
with which expression, to how high order, and for how many number of
flavors.

In Eq.\ (\ref{agenthree}), there are other two arbitrary constants
$\Lambda$ and $C$. Because the RG equation (\ref{RGa}) or (\ref{aprga})
is of the first order, only one of them is independent. So we can
arbitrarily take one of them, while the other one is determined by
giving an initial condition.
Although the freedom $C$ can be useful, e.g., for an optimization
of the finite-order perturbation calculations and construction of
universal $N_{\mathrm{f}}$-independent $\Lambda$ \cite{Marciano1984PRD29} etc,
it nearly becomes standard, nowadays,
to take $C=0$ \cite{BardeenPRD18,BurasRMP52}, which makes
the expression simpler (one would fail in Eq.\ (\ref{alfaprox1})
to determine $C$ for a given $\Lambda$\ value because Eq.\ (\ref{alfaprox1})
has nothing to do with $C$), and \cite{PDG2020}
\begin{equation} \label{aini}
\alpha_N^{(i)}(M_{\mathrm{Z}},N_\mathrm{f},\Lambda,0)=0.1179/\pi,
\end{equation}
where $M_{\mathrm{Z}}=91.1876$ GeV
is the mass of Z bosons.
With the initial condition in Eq.\ (\ref{aini}),
we can determine the scale $\Lambda$.

Please note, the experimental average of $\alpha_s(M_{\mathrm{Z}})$
has moved by less than 1$\sigma$\ from the previous one in Ref.\
\cite{Eidelman2004PLB592.1}.

From Eq.\ (\ref{aini}) we know that the scale $\Lambda$\ depends
on $N_{\mathrm{f}}, N, i$, i.e.,
\begin{equation}
\Lambda=\Lambda_n(N,i).
\end{equation}
Here $n=N_{\mathrm{f}}=3,4,5,6$, i.e.,
setting $C=0$ requires distinct $\Lambda$\ for
different effective flavor regimes,
and we use $\Lambda_3$,  $\Lambda_4$,
$\Lambda_5$, and $\Lambda_6$ for $m_s<u<m_c$, $m_c<u<m_b$,
$m_b<u<m_t$, and $u>m_t$, respectively, where
the relevant quark thresholds are taken, in the present calculations,
to be $m_t=172.76$ GeV, $m_b=4.18$ GeV, $m_c=1.27$ GeV, and $m_s=93$ MeV
\cite{PDG2020}.

\begin{figure}[htb]
\centering
\includegraphics[width=8.2cm]{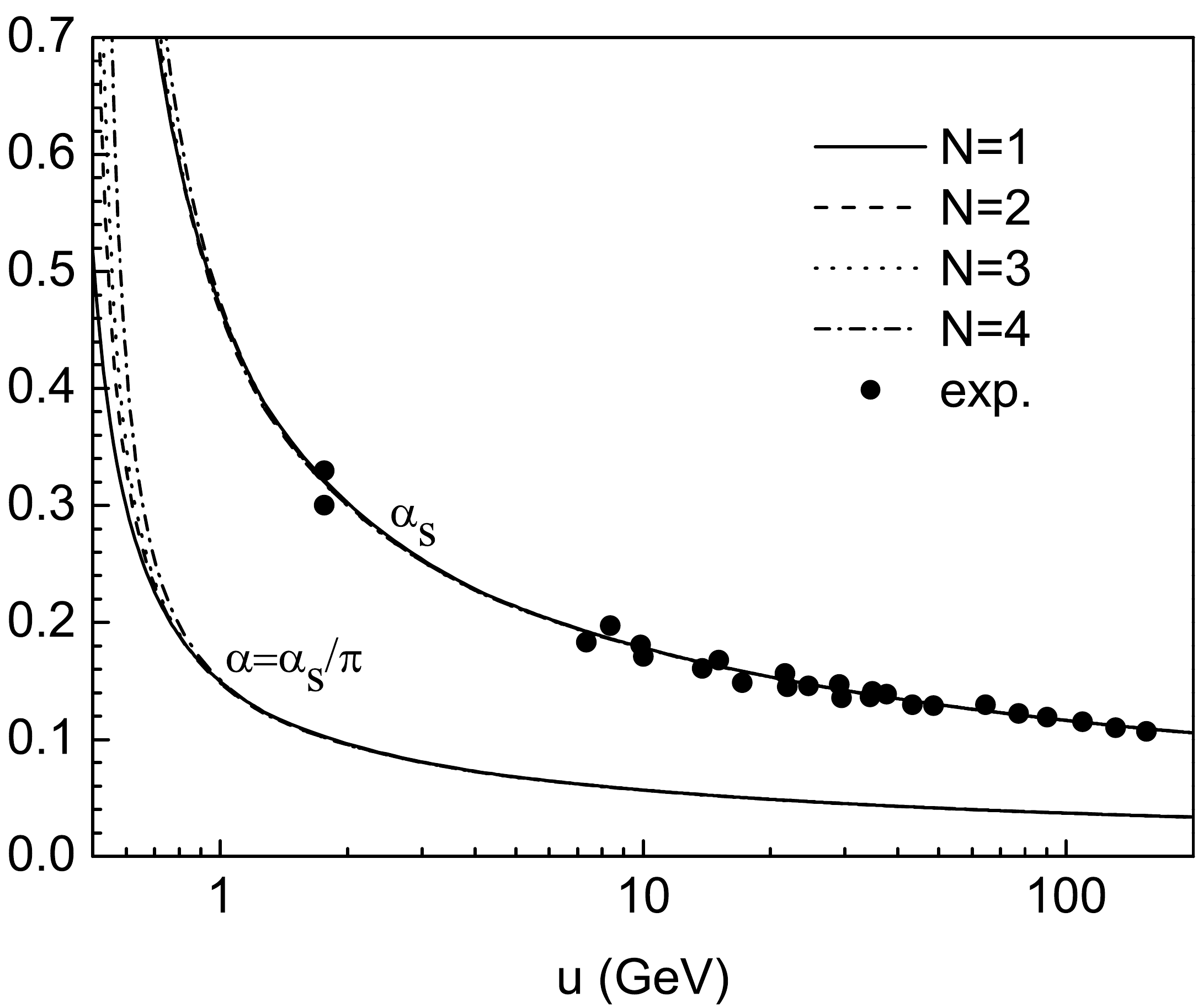}
\caption{
 The QCD coupling as a function of the 't Hooft unit of mass $u$,
 calculated by the new expression in Eq.\ (\ref{alfgen2})
 for the order $N$ from 1 to 4. 
 in Tab.\ \ref{tabLam}. The full dots are the experimental data
 of, in increasing  $u$,  $\tau$\ decay, low $\mathrm{Q^2}$ cont, and $\mathrm{EW}$ precision fit
  from the $\mathrm{N^3LO}$ data, e$^+$e$^-$ and $\mathrm{pp/p\bar{p}}$ event shapes from the $\mathrm{NNLO}$ data,
 Heavy Quarkonia, $\mathrm{DIS}$ jets, and $\mathrm{pp}$ event shape from the $\mathrm{NLO}$ data.
         }
 \label{aL}
\end{figure}

\begin{figure}[htb]
\centering
\includegraphics[width=8.2cm]{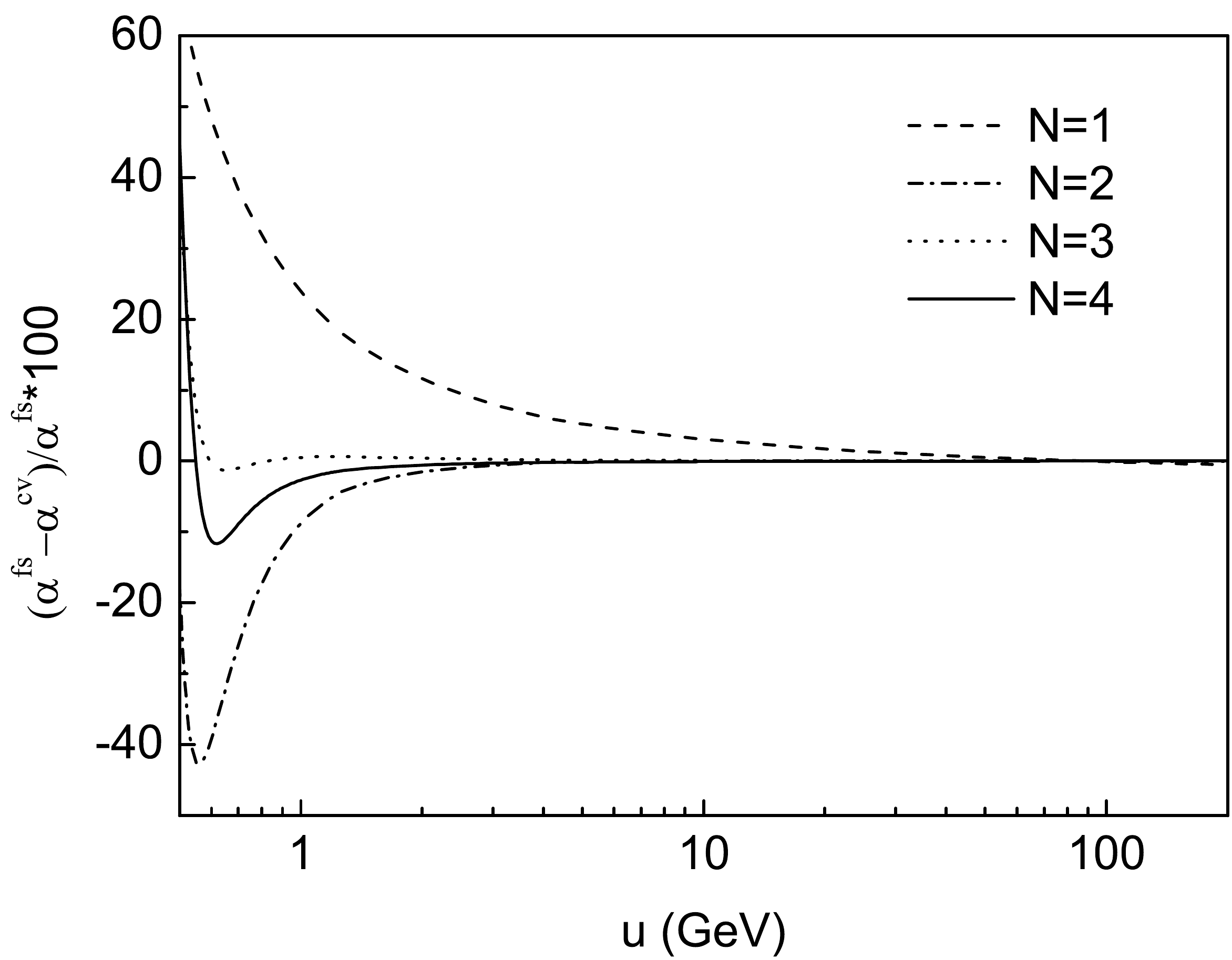}
\caption{
 The relative difference between the new expansion in Eq.\ (\ref{alfgen2})
 and the conventional expansion in Eq.\ (\ref{alfexpn}).
         }
 \label{aL2}
\end{figure}

In Fig.\ \ref{aL}, the coupling is shown as a function of the
renormalization point, calculated from Eq.\ (\ref{alfgen3})
with the order $N$ from 1 to 4.
Please note, the beta coefficients used in the calculation are,
in fact, the primed beta coefficients in Eq.\ (\ref{primedbeta}),
rather than the original ones in Eq.\ (\ref{originalbeta}).
Therefore, the corresponding coupling is mathching-invariant,
to be discussed in detail a little later.
The same calculation has also been performed from the conventional
expansion in Eq.\ (\ref{alfappr01}). The relative difference between the
results from Eq.\ (\ref{alfgen3}) and Eq.\ (\ref{alfappr01}) is shown in
Fig.\ \ref{aL2}. It can be seen that, with decreasing $u$, the difference
becomes more and more significant.

\begin{table}
\centering
\caption{
\label{tabLam}
 The QCD scale parameter $\Lambda$\ for the order $N$ from 1 to 4
 and the number of flavors from 3 to 6.
 There are three rows for each $\Lambda_n$:
 the first row is calculated from the new expression in Eq.\ (\ref{alfgen3}),
 the second row from the conventional standard expansion in Eq.\ (\ref{alfappr01}),
 and the third row from Eq.\ (\ref{aexp}).
        }
\begin{tabular}{ccccccccc}\hline
$\Lambda$\mbox{(MeV)} && $N=1$ && $N=2$ && $N=3$ && $N=4$ \\
\hline
           &&314.987 &&336.766 &&330.620 &&329.181 \\
$\Lambda_3$&&141.792 &&367.144 &&324.305 &&327.787 \\
           &&141.788 &&355.335 &&328.907 &&329.829 \\
 \hline
           &&275.009 &&291.038 &&288.796 &&288.129 \\
$\Lambda_4$&&118.981 &&321.668 &&285.921 &&287.621 \\
           &&118.983 &&300.259 &&285.120 &&288.258 \\
 \hline
           &&199.080 &&207.159 &&207.732 &&207.450 \\
$\Lambda_5$&&87.3112 &&224.966 &&207.199 &&207.199 \\
           &&87.3112 &&210.699 &&204.334 &&207.464 \\
  \hline
           &&84.2465 &&86.4323 &&87.4815 &&87.3847 \\
$\Lambda_6$&&42.3769 &&90.9752 &&87.6606 &&87.2738 \\
           &&42.3860 &&87.3419 &&85.8268 &&87.4247 \\
  \hline
\end{tabular}
\end{table}

\begin{figure}[thb]
\centering
\includegraphics[width=8.2cm]{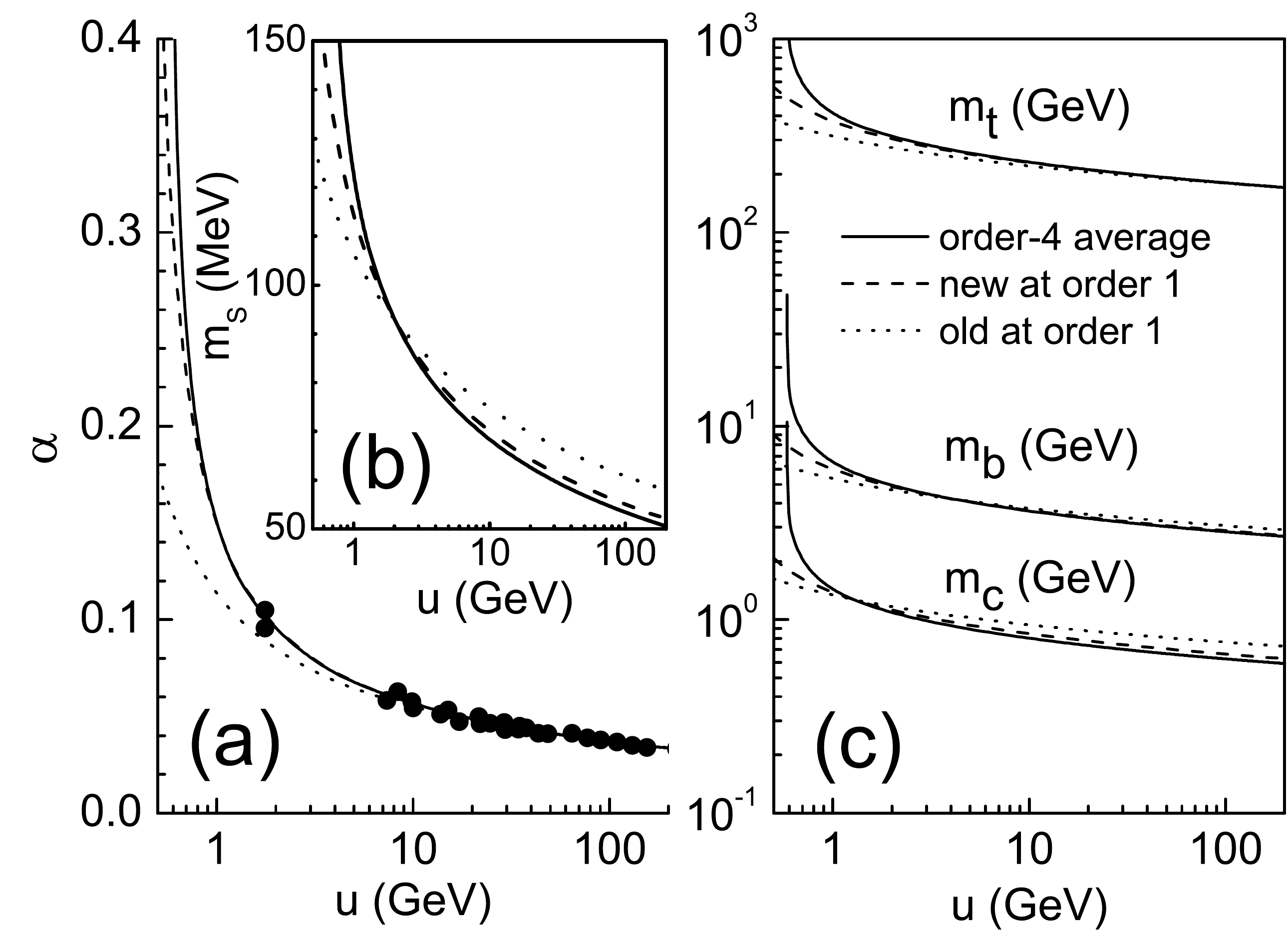}
\caption{
\label{figam}
  QCD running coupling and quark masses in different expressions.
The full line is the average value of the new and previous approaches
at order 4, the dashed and dotted lines are, respectively, from the new expressions in Eq.\ (\ref{alfgen3})
and previous expressions in Eq.\ (\ref{aexp}), at leading order. The new expression
converges much faster.
        }
\end{figure}

To compare convergence speed of three expressions in Eqs.\ (\ref{alfgen3}),
(\ref{alfappr01}), and (\ref{aexp}),
 all the $\Lambda_n(N,i)$, where
$n=3$---6, $N=1$---4, and $i=$ cv/rt/fs,
are listed in Tab.\ \ref{tabLam}.
There are three rows corresponding to each $\Lambda_n$:
the first row is for Eq.\ (\ref{alfgen3}),
the second row for Eq.\ (\ref{alfappr01}),
and the third row for Eq.\ (\ref{aexp}).
It is obvious that the new expression in Eq.\ (\ref{alfgen3})
converges much faster than the previous ones.
Even at the leading-order ($N=1$), the corresponding $\Lambda_n$
from Eq.\ (\ref{alfgen3}) has already reached less than 5\% close to the value
at order 4, while the other two have still been 50\% away from the
order-4 value.
The rapid convergent speed of the new expression in Eq.\ (\ref{alfgen3})
 can also be directly seen from Fig.\ \ref{figam}(a), where
the leading order coupling from the new expression (dashed line)
is compared with the one from the previous expression (dotted line)
and the order-4 average of the three expressions (solid line).

It is easy to understand the faster convergence of $\alpha^{\mathrm{fs}}$
than $\alpha^{\mathrm{cv}}$ due to the inclusion of an infinite number of
known logarithmic terms in a closed form. To understand that
$\alpha^{\mathrm{fs}}$ converges faster than $\alpha^{\mathrm{rt}}$,
we can expand $\alpha_N^{\mathrm{rt}}$ to a series which is nothing
but exactly the same as that in Eq.\ (\ref{ageneral}) with
all $\beta_{i\ge N}$ set to zero. Therefore, $\alpha^{\mathrm{rt}}$
contains more terms than $\alpha^{\mathrm{cv}}$ and $\alpha^{\mathrm{fs}}$.
However, the extra terms, which are included in $\alpha^{\mathrm{rt}}$
but not in $\alpha^{\mathrm{fs}}$,
are incorrect because all $\beta_{i\ge N}$ have been set to zero.

\section{Matching-invariant quark masses}
\label{secqkmass}

The renormalization equation for the quark masses is
\begin{equation} \label{rgmq}
\frac{\mathrm{d}\ln m_{\mathrm{q}}}{\mathrm{d}\ln u^2}
=-\sum_{i=0}^{\infty} \gamma_i\alpha^{i+1}
\equiv \gamma(\alpha).
\end{equation}
where the quark mass anomalous dimension $\gamma_i$
is known to 4-loop level \cite{Chetyrkin1997PLB404,Vermaseren1997PLB405}:
\begin{subequations}
 \label{rexp}
\begin{eqnarray}
\gamma_0&=&1,   \label{rexp01}
\ \
\gamma_1=101/24-(5/36) N_{\mathrm{f}}, \\
\gamma_2
&=&
  \frac{1249}{64}
  -\left(
    \frac{5}{6}\zeta_3+\frac{277}{216}
   \right)
   N_{\mathrm{f}}
  -\frac{35}{1296}N_{\mathrm{f}}^2,  \\
\gamma_3
&=&
  \frac{4603055}{41472}
  +\frac{530}{27}\zeta_3
  -\frac{275}{8}\zeta_5
    \nonumber\\
&&
 -\left(
   \frac{91723}{6912}
   +\frac{2137}{144}\zeta_3
   -\frac{575}{72}\zeta_5
   -\frac{11}{288}\pi^4
  \right)
  N_{\mathrm{f}}
    \nonumber\\
&&
 +\left(
   \frac{2621}{31104}
   +\frac{25}{72}\zeta_3
   -\frac{\pi^4}{432}
  \right)
  N_{\mathrm{f}}^2
    \nonumber\\
&&  \label{rexpe}
 +\left(
   \frac{\zeta_3}{108}
   -\frac{83}{15552}
  \right)
  N_{\mathrm{f}}^3.
\end{eqnarray}
\end{subequations}

Suppose the matching condition for quark masses is
\begin{equation} \label{mcmq}
m_q^*
=m_q\sum_{i=0}^{\infty}
    \sum_{j=0}^i
    D_{i,j}\alpha^i\ln^j\frac{m_{\mathrm{\mathrm{h}}}^2}{u^2},
\end{equation}
i.e.,
$m_q$ is the quark mass in the full
$N_{\mathrm{f}}$-flavor theory for flavor $q$, while $m_q^*$ belongs to
the corresponding effective theory with $N_{\mathrm{f}}-1$ flavors.
$m_q^*$ should also satisfy RG equation, and thus satisfy
\begin{equation} \label{mqseq}
\frac{\mathrm{d}m_q^*}{m_q\mathrm{d}\ln u^2}
=-\frac{m_q^*}{m_q}\sum_{k=0}^{\infty} \gamma_k^*{\alpha^*}^{k+2}.
\end{equation}
Substituting Eq.\ (\ref{mcmq}) into Eq,\ (\ref{mqseq}), and then
comparing coefficients will give $D_{i,j>0}$ recursively
%
\begin{eqnarray}
D_{i,j}
&=&\frac{1}{j}
  \sum_{i'=0}^{\bar{\imath}}\sum_{j'=J^*}^{\bar{\jmath}}
  D_{\bar{\imath}-i',\bar{\jmath}-j'}
    \sum_{k=0}^{i'-j'}\gamma_k^*\bigsqcup_{0,0}^{k+1}C_{i'-k,j'}
 \nonumber\\
&&
   -\frac{1}{j}\sum_{k=\bar{\jmath}}^{\bar{\imath}}
     \left(
      \gamma_{k-\bar{\jmath}}D_{\bar{\imath}+\bar{\jmath}-k,\bar{\jmath}}
      +k \beta_{\bar{\imath}-k} D_{k,\bar{\jmath}}
     \right)
\nonumber\\
&&
 -2\sum_{k=j}^{i-1}\gamma_{i-1-k}D_{k,j},
 \label{Dijeq}
\end{eqnarray}
where $J^*=\mathrm{max}(0,j-i+i'),\ \bar{\imath}=i-1,\ \bar{\jmath}=j-1$.

To solve $D_{i,j>0}$ from Eq.\ (\ref{Dijeq}), we should know
$D_{i,0}$. Here are the results for $D_{i,0}$ to three-loop level
\cite{Bernreuther1982NPB197,Larin1995NPB438}:
\begin{subequations} \label{Di0val}
\begin{eqnarray}
D_{0,0}&=&1, \ \ D_{1,0}=0, \ \ D_{2,0} = 89/432, \\
D_{3,0}
&=&
  \left(
   \frac{1327}{11664}
   -\frac{2}{27}\zeta_3
  \right) N_{\mathrm{f}}
  +D_{3,0}^{(0)}.
\end{eqnarray}
\end{subequations}
With the quadrilogarithm
$
\mathrm{Li}_4(1/2)
=\mbox{polylog}(4,1/2)
=\sum_{k=1}^{\infty}
 (1/2)^k/k^4
\approx 0.5174790617,
$
the constant $D_{3,0}^{(0)}$ is
$
D_{3,0}^{(0)}
\equiv
 {10477}/{11664}
 -({343}/{864})\zeta_3
 +\frac{103}{6480}\pi^4
 -\frac{4}{9}\mbox{Li}_4(1/2)
 +\frac{\ln^2 2}{54}\left(\pi^2-\ln^2 2\right)
\approx 1.822899133
$\
\cite{Chetyrkin1998npb510}.

Solving Eq.\ (\ref{Dijeq}) with the aid of Eq.\ (\ref{Di0val}) gives
\begin{subequations}
\begin{eqnarray}
D_{1,1}
&=&
  0, \ \ D_{2,1}=5/18, \ \ D_{2,2}=1/3, \\
D_{3,1}
&=&
   D_{3,1}^{(0)}
  +\frac{53}{216} N_{\mathrm{f}}
  ,\
D_{3,2}
= \frac{175}{108},
  \\
D_{3,3}
 &=& \frac{1}{27}
     \left( 2 N_{\mathrm{f}} -31 \right),\
D_{4,1}
=
  D_{4,1}^{(0)}
  +D_{4,1}^{(1)}N_{\mathrm{f}}
  +D_{4,1}^{(2)}N_{\mathrm{f}}^2,
 \nonumber\\
D_{4,2}
&=&
   D_{4,2}^{(0)}
  -\frac{8321}{2592} N_{\mathrm{f}}
  +\frac{31}{324} N_{\mathrm{f}}^2,\
D_{4,3}
=
  \frac{23}{36} N_{\mathrm{f}}
  -\frac{2615}{324},
  \nonumber \\
D_{4,4}
&=&
  \frac{1039}{216}
  -\frac{16}{27} N_{\mathrm{f}}
  +\frac{N_{\mathrm{f}}^2}{54},\ \cdots
\end{eqnarray}
\end{subequations}
where
$
D_{3,1}^{(0)}
=(5/3)\zeta_3 -7/1296
\approx 1.998026937,
\
D_{4,1}^{(0)}
= 644891/20736
 -(33/2)D_{3,0}^{(0)}
 -35\pi^4/432
 +97175\zeta_3/4608
 -575\zeta_5/36
\approx 1.91761725,
\
D_{4,1}^{(1)}
= D_{3,0}^{(0)}
 +\pi^4/108
 -\zeta_3/9
 -35173/15552
\approx 0.3296349104,
\
D_{4,1}^{(2)}
=(3401/432-7\zeta_3)/54
\approx -0.0100317247,
\
D_{4,2}^{(0)}
= 14809/648 -155\zeta_3/12
\approx 7.32682673.
$

If the $m_{\mathrm{h}}$ in Eq.\ (\ref{mcmq}) is RG-invariant,
the last term on the right hand side of Eq.\ (\ref{Dijeq})
disappear.
  On the other hand, if one takes $u=m_{\mathrm{h}}$,
then the matching condition Eqs.\ (\ref{mcmq}) becomes
\begin{eqnarray}
m_q^*
&=&
 m_q
 \sum_{i=0}^{\infty}
 D_i \alpha^i,
 \label{mqstar0}
\end{eqnarray}
where $D_i\equiv D_{i,0}$. The matching condition for the running coupling is
\begin{eqnarray}
\alpha^*
&=&
 \sum_{k=0}^{\infty}
 C_k\alpha^{k+1},
 \label{alphastar0}
\end{eqnarray}
where $C_k\equiv C_{k,0}$ is written in the appendix \ref{mat-alf}.

Suppose the matching-invariant quark masses are given by
\begin{equation} \label{newmdef}
m_q'
=m_q
 \sum_{j=0}^{\infty}
 b_j \alpha^j,
\end{equation}
then the matching condition for $m_q'$ is
\begin{eqnarray}
{m_q'}^*
=m_q^*
 \sum_{j=0}^{\infty} b_j^*{\alpha^*}^j,
\label{mqps}
\end{eqnarray}
where $b_i$ depends on $N_{\mathrm{f}}$
and is to be determined by the condition
${m_{\mathrm{q}}'}^*=m_q'$. Again a superscript star means
decreasing $N_{\mathrm{f}}$ by one.

Substituting Eqs.\ (\ref{mqstar0}) and (\ref{alphastar0})
into Eq.\ (\ref{mqps}), and then comparing the coefficients
of $\alpha^i$\ in the equality ${m_{\mathrm{q}}'}^*=m_q'$,
we obtain
\begin{equation} \label{bibip}
b_i=\sum_{k=0}^i D_{i-k} \sum_{j=0}^kb_j^*\bigsqcup_0^j C_{k-j}.
\end{equation}

Express $b_i$ as a polynomial of $N_{\mathrm{f}}$, i.e.,
\begin{equation}
b_i=\sum_{k=0}^i b_{i,k} N_{\mathrm{f}}^i,
\end{equation}
then $b_j^*=\sum_{l=0}^j b_{j,l}(N_{\mathrm{f}}-1)^l$,
and Eq.\ (\ref{bibip}) becomes
\begin{equation}
\sum_{k=0}^i
\left[
 b_{i,k}N_{\mathrm{f}}^k
 -D_{i-k}
  \sum_{j=0}^k
  \sum_{l=0}^j b_{j,l} (N_{\mathrm{f}}-1)^l
   \bigsqcup_0^j C_{k-j}
\right]
=0
\end{equation}
whose solution is
\begin{subequations}
\begin{eqnarray}
b_{1,1}&=& 0, \ \
b_{2,1} = (89/432)b_{0,0}, \ \ b_{2,2}=0, \\
b_{3,1}&=&
  \left(
     \frac{7427}{7776}
     -\frac{125}{288}\zeta_3
     +\frac{\pi^4}{72}
     -\frac{B_4}{36}
    \right) b_{0,0}
 \nonumber\\
&&
    +\frac{155}{432}b_{1,0} \\
 b_{3,2}
&=&
   \left(
     \frac{1327}{23328}
     -\frac{\zeta_3}{27}
    \right)
    b_{0,0},
 \ \ b_{3,3}=0, \\
&\cdots&  \nonumber
\end{eqnarray}
\end{subequations}
where
$
B_4=16\mbox{Li}_4(1/2)-13\zeta_4/2-4\zeta_2\ln^22+(2/3)\ln^42
\approx -1.762800087
$ \cite{Chetyrkin1998npb510}.
Taking the simplest nontrivial choice $b_{i,0}=\delta_{i,0}$ leads to
\begin{equation} \label{bval}
b_0=1,\ \ b_1=0,\ b_2=\frac{89}{432}N_{\mathrm{f}},\
b_3= b_{3,1}
  N_{\mathrm{f}}
    +b_{3,2} N_{\mathrm{f}}^2,\ \cdots
\end{equation}
where
\begin{subequations} \label{b31b32}
\begin{eqnarray}
b_{3,1}
&=& \frac{7427}{7776}
  -\frac{125}{288}\zeta_3
  +\frac{\pi^4}{72}
  -\frac{B_4}{36}
  \nonumber\\
&\approx&
 1.835262938, \\
b_{3,2}
&=& \frac{1327}{23328}-\frac{\zeta_3}{27}
\approx 0.01236380468.
\end{eqnarray}
\end{subequations}
Explicitly, the matching-invariant quark mass is
\begin{equation}
\frac{m_q'}{m_q}
=
  1+\frac{89}{432}N_{\mathrm{f}} \alpha^2
  +\left[
    b_{3,1}+b_{3,2} N_{\mathrm{f}}
   \right]
   N_{\mathrm{f}} \alpha^3
  +\cdots.
\end{equation}

The inverse relation of Eq.\ (\ref{newmdef}) is
\begin{equation}
m_q
=m_q'
 \sum_{j=0}^{\infty}b_j'{\alpha'}^j,
\end{equation}
where the coefficients $b_j'$ are linked to $b_j$ by solving
\begin{equation}
\sum_{i=0}^k b_{k-i}
\sum_{j=0}^ib_j' \bigsqcup_0^j a_{i-j} = \delta_{k,0}.
\end{equation}
To three-loop level, one has the explicit expression
\begin{eqnarray}
\frac{m_q}{m_q'}
  =
  1-\frac{89}{432}N_{\mathrm{f}} {\alpha'}^2
   -\left[
 b_{3,1}+b_{3,2} N_{\mathrm{f}}
   \right]
   N_{\mathrm{f}} {\alpha'}^3
  +\cdots,
\end{eqnarray}
where $b_{3,1}$ and $b_{3,2}$ are given in Eq.\ (\ref{b31b32}).

The RG equation for the new mass in the new coupling is
\begin{equation}
\frac{\mathrm{d}\ln m_q'}{\mathrm{d}\ln u^2}
=-\sum_{i=0}^{\infty} \gamma_i' {\alpha'}^{i+1}
\end{equation}
where $\gamma'$ is the new gamma coefficients
determined by
\begin{equation}
\sum_{k=0}^i
b_{i-k}
\left[
 \sum_{j=0}^k
  \gamma_j'
  \bigsqcup_0^{j+1} a_{k-j}
  -\gamma_k
  -(i-k)\beta_k
\right]
=0,
\end{equation}
which is also a recursive relation. The solution is
\begin{subequations}
\begin{eqnarray}
\gamma_0' &=& \gamma_0,\ \gamma_1'=\gamma_1,\
\gamma_2' = \gamma_2-a_2\gamma_0+2b_2\beta_0, \\
\gamma_3'
&=&
   \gamma_3
  -2a_2\gamma_1-a_3\gamma_0+2b_2\beta_1+3b_3\beta_0,\ \cdots
\end{eqnarray}
\end{subequations}
Here $a_{0}=b_0=1$ and $a_1=b_1=0$ have been used.
With the expressions for $a_2$ and $a_3$ in Eq.\ (\ref{aval}),
$b_2$ and $b_3$ in Eq.\ (\ref{bval}), the original beta and gamma
coefficients in Eqs.\ (\ref{originalbeta}) and (\ref{rexp}), the primed
gamma coefficients can be expressed as
\begin{equation}
\gamma_i'=\sum_{k=0}^i c_{i,k} N_{\mathrm{f}}^k
\end{equation}
with the color factors given by
$
c_{0,0}=1,\ c_{1,0}={101}/{24},\ c_{1,1}=-{5}/{36},\
c_{2,0}={1249}/{64},\
c_{2,1}=-{29}/{96}-{5}\zeta_3/6\approx -1.303797419,\
c_{2,2}=79/1152,\
\gamma_{3,0}
 = 4603055/41472 +530\zeta_3/27 -275\zeta_5/8
  \approx 98.94341426, \
\gamma_{3,1}
 = -357941/41472
   -427261\zeta_3/27648
   +575\zeta_5/72
   +11\pi^4/72
   -11B_4/48
  \approx -3.640068635, \
\gamma_{3,2}
 = -10291/62208 +149\zeta_3/576
   -\pi^4/108 +B_4/72
   \approx -0.7808994989, \
\gamma_{3,3}
 = \zeta_3/36 -197/5832 \approx -0.0003886799877.
$

%

\section{expression for quark masses}
\label{exp-qm}

To solve Eq.\ (\ref{rgmq}), one can divide it by
Eq.\ (\ref{RGa}), giving
\begin{equation} \label{rgmq2}
\frac{d\ln m_q}{d\alpha}
=\frac{\gamma(\alpha)}{\beta(\alpha)}
= \frac{\gamma_0}{\beta_0\alpha}
  +\acute{\gamma}(\alpha),
\end{equation}
where
\begin{equation}
\acute{\gamma}(\alpha)
\equiv
  \frac{\gamma(\alpha)}{\beta(\alpha)}
 -\frac{\gamma_0}{\beta_0\alpha}
=\frac{
  \sum_{i=0}^{\infty}
  (\gamma_{i+1}-\frac{\gamma_0}{\beta_0}\beta_{i+1})
   \alpha^i}
      {\sum_{j=0}^{\infty}\beta_j\alpha^j}.
\end{equation}
It is easy to check that $\acute{\gamma}(\alpha)$ is analytic
at $\alpha=0$, and can thus be expanded to a Taylor series, i.e.,
\begin{equation} \label{agamtalor}
\acute{\gamma}(\alpha)
 =\sum_{k=0}^{\infty} \acute{\gamma}_k\alpha^k,
\end{equation}
where
the Taylor coefficients can be obtained
either from the normal mathematical formula
\begin{equation}
\acute{\gamma}_k
=\frac{1}{k!} \left.
 \frac{\mathrm{d}^k}{\mathrm{d}\alpha^k}\acute{\gamma}(\alpha)
              \right|_{\alpha=0},
\end{equation}
or, simply, from the recursive relation
\begin{equation} \label{Bipp}
\acute{\gamma}_k
= \frac{\gamma_{k+1}}{\beta_0}
 -\frac{\gamma_0}{\beta_0^2}\beta_{k+1}
 -\frac{1}{\beta_0}\sum_{l=0}^{k-1} \beta_{k-l} \acute{\gamma}_l.
\end{equation}
The following are the first several concrete expressions:
\begin{subequations}
\begin{eqnarray}
\acute{\gamma}_0
&=&
 \gamma_1/\beta_0-\gamma_0\beta_1/\beta_0^2,\\
\acute{\gamma}_1
&=&
 {\gamma_2}/{\beta_0}
 -(\beta_1\gamma_1+\beta_2\gamma_0)/{\beta_0^2}
 +{\beta_1^2\gamma_0}/{\beta_0^3},\\
\acute{\gamma}_2
&=&
 {\gamma_3}/{\beta_0}
 -(\beta_1\gamma_2+\beta_2\gamma_1+\beta_3\gamma_0)/\beta_0^2
\nonumber\\
&&
 +\beta_1(\beta_1\gamma_1+2\beta_2\gamma_0)/\beta_0^3
  -\beta_1^3\gamma_0/\beta_0^4,\\
&\cdots& \nonumber
\end{eqnarray}
\end{subequations}

Integrating Eq.\ (\ref{rgmq2}) leads to
\begin{equation} \label{intqkm}
\ln\frac{m_q}{\hat{m}_q}
=\frac{\gamma_0}{\beta_0}\ln\alpha
 +\int_0^{\alpha}
  \acute{\gamma}(x)
  \mathrm{d} x,
\end{equation}
where $\hat{m}_q$ is a RG-invariant dimensional parameter, or
the constant of integration.
Eq.\ (\ref{intqkm}) immediately gives
\begin{equation}  \label{mqexp}
m_q
=\hat{m}_q \alpha^{{\gamma_0}/{\beta_0}}
 \exp\left[
 \int_0^{\alpha}\acute{\gamma}(x)
 \mathrm{d} x
     \right].
\end{equation}
One can expand the last factor on the right hand side
of Eq.\ (\ref{mqexp}) to a Taylor series, giving
\begin{equation} \label{mqexpd01}
m_q
=\hat{m}_q \alpha^{{\gamma_0}/{\beta_0}}
 \left(
  \sum_{i=0}^{\infty} \tilde{\gamma}_i \alpha^i
 \right)
\end{equation}
where
\begin{equation}
\tilde{\gamma}_i
=\frac{1}{i!}
 \left.
  \frac{d^i}{d\alpha^i}
    \exp\left[\int_0^{\alpha} \acute{\gamma}(x)\mathrm{d}x\right]
 \right|_{\alpha=0}.
\end{equation}
In fact, substituting Eq.\ (\ref{agamtalor}) into Eq.\ (\ref{mqexp}),
integrating term by term, aplying $e^x=\sum_{k=0}^{\infty}x^k/k!$,
and then using the square-cup operator,
we can give a general expression
\begin{equation}
\tilde{\gamma}_i
=\sum_{k=0}^i
  \frac{1}{(i-k)!}
  \bigsqcup_0^{i-k} \frac{\acute{\gamma}_k}{k+1}.
\end{equation}
Explicitly, the tilde gamma can be listed as
\begin{subequations}
\begin{eqnarray}
\tilde{\gamma}_0
 &=& 1,\
\tilde{\gamma}_1
 =\acute{\gamma}_0
 ={\gamma_1}/{\beta_0}
  -{\gamma_0\beta_1}/{\beta_0^2}, \\
\tilde{\gamma}_2
&=&
  \left(\acute{\gamma}_0^2+\acute{\gamma}_1\right)/2
 \nonumber\\
&=&
  \frac{1}{2}
  \left[
   {\gamma_2}/{\beta_0}
   +({\gamma_1^2-\gamma_1\beta_1-\gamma_0\beta_2})/{\beta_0^2}
  \right.\nonumber\\
&&\left.\phantom{\frac{1}{2}[}
   -{\gamma_0\beta_1(2\gamma_1-\beta_1)}/{\beta_0^3}
   +{\gamma_0^2\beta_1^2}/{\beta_0^4}
  \right], \\
\tilde{\gamma}_3
&=&
 \frac{1}{6}\acute{\gamma}_0^3
 +\frac{1}{2}\acute{\gamma}_0\acute{\gamma}_1
 +\frac{1}{3}\acute{\gamma}_2
\nonumber\\
&=&
  \frac{\gamma_3}{3\beta_0}
  +\frac{3\gamma_1\gamma_2
         -2(\gamma_0\beta_3+\gamma_1\beta_2+\gamma_2\beta_1)
        }
        {6\beta_0^2}
   \nonumber\\
&& \hspace{-0.8cm}
 +\frac{\gamma_1^3
       +4\gamma_0\beta_1\beta_2
       +2\gamma_1\beta_1^2
       -3(\gamma_1^2\beta_1
       +\gamma_0\gamma_2\beta_1
       +\gamma_0\gamma_1\beta_2)
       }
       {6\beta_0^3}
   \nonumber\\
&&
 +\frac{\gamma_0\beta_1
       (3\gamma_0\beta_2
        +6\gamma_1\beta_1
        -2\beta_1^2-3\gamma_1^2
       )}
      {6\beta_0^4}
  \nonumber\\
&&
 +\frac{\gamma_0^2\beta_1^2(\gamma_1-\beta_1)}
   {2\beta_0^5}
 -\frac{\gamma_0^3\beta_1^3}{6\beta_0^6}, \\
&\cdots& \nonumber
\end{eqnarray}
\end{subequations}

To have a finite-form expression for quark masses, one can
truncate the series in Eq.\ (\ref{mqexpd01}) to a finite order,
say $N$, to give
\begin{equation} \label{mqexpdfin}
m_q
=\hat{m}_q \alpha^{{\gamma_0}/{\beta_0}}
 \left(
  \sum_{i=0}^{N-1} \tilde{\gamma}_i \alpha^i
 \right).
\end{equation}

Another approach is to truncate the series $\acute{\gamma}(x)$
in Eq.\ (\ref{mqexp}). In this way
the solution in Eq.\ (\ref{mqexp}) becomes
\begin{equation} \label{mqexpfinexact}
m_q
=\hat{m}_q \alpha^{{\gamma_0}/{\beta_0}}
 \exp\left[
       w_N(\alpha)
     \right].
\end{equation}
where
\begin{equation}
w_N(\alpha)
\equiv
\int_0^{\alpha}
  \frac{\sum_{j=0}^{N-2}
   (\gamma_{j+1}-\frac{\gamma_0}{\beta_0}\beta_{j+1})x^j}
       {\sum_{i=0}^{N-1}\beta_i x^i}
 \mathrm{d} x.
\end{equation}
Explicitly, one can derive
\begin{subequations}
\begin{eqnarray} \hspace{-0.5cm}
w_1(\alpha)
&=&0, \\
w_2(\alpha)
&=&\left(
  \frac{\gamma_1}{\beta_1}-\frac{\gamma_0}{\beta_0}
 \right)
 \ln\left(
     1+\frac{\beta_1}{\beta_0}\alpha
    \right), \\
w_3(\alpha)
&=&\frac{1}{2}
  \left(
  \frac{\gamma_2}{\beta_2}-\frac{\gamma_0}{\beta_0}
 \right)
 \ln\left(
     1+\frac{\beta_1}{\beta_0}\alpha
      +\frac{\beta_2}{\beta_0}\alpha^2
    \right) \nonumber\\
&& \hspace{-0.8cm}
  -\frac{
    \left(
      \frac{\gamma_0}{\beta_0}
     -2\frac{\gamma_1}{\beta_1}
     +\frac{\gamma_2}{\beta_2}
    \right)
        }
        {
     \sqrt{4\beta_0\beta_2/\beta_1^2-1}
        }
    \left[
    \mbox{arctan}\frac{1+2\frac{\beta_2}{\beta_1}\alpha}
                      {\sqrt{4\beta_0\beta_2/\beta_1^2-1}}
    \right.
 \nonumber\\
&& \left.
   -\mbox{arctan}\frac{1}{\sqrt{4\beta_0\beta_2/\beta_1^2-1}}
   \right],
  \\
w_4(\alpha)
&=&
 \iota_4
 \left[
  \arctan\frac{\alpha-\iota_1}{\sqrt{\iota_0-\iota_1^2}}
  +\arctan\frac{\iota_1}{\sqrt{\iota_0-\iota_1^2}}
 \right]
\nonumber\\
&&
 +\ln\frac{\left(
            1-2\iota_1\alpha/\iota_0+\alpha^2/\iota_0
           \right)^{\iota_3}}
        {\left(1+\alpha/\alpha_0\right)^{\iota_2}},
\end{eqnarray}
\end{subequations}
where
%
\begin{eqnarray}
\iota_0
&=&
 \frac{\beta_0}{\beta_3\alpha_0}, \
\iota_1
 =\frac{1}{2}\left( \alpha_0-\frac{\beta_2}{\beta_3} \right), \\
\iota_2
&=&
 \frac{
  \sum_{i=1}^3 (-1)^i
  \left(\gamma_i-\frac{\gamma_0}{\beta_0}\beta_i\right)\alpha_0^{i-1}
      }
      {
  \beta_2\alpha_0-2\beta_1+3\beta_0/\alpha_0
      }, \\
\iota_3
&=&
 \frac{1}{2}
 \left(
  \iota_2+\frac{\gamma_3}{\beta_3}-\frac{\gamma_0}{\beta_0}
 \right), \\
\iota_4
&=&
 \frac{
   (\gamma_1/\beta_3-\gamma_0\beta_1/\beta_0\beta_3
        +\iota_0\iota_2)/\alpha_0
  +2\iota_1\iota_3
    }
    {\sqrt{\iota_0-\iota_1^2}}.
\end{eqnarray}

Previous calculations usually use Eq.\ (\ref{mqexpdfin})
or (\ref{mqexpfinexact}) to calculate quark running masses.
The weak point is obvious: the dependence on $u$ is not direct, but
through $\alpha$ whose dependence on $u$ is not exactly known, which
causes extra imprecision. To connect $m_q$ directly to $u$,
we can substitute the general expression for $\alpha$\
in Eq.\ (\ref{ageneral}) into Eq.\ (\ref{mqexpd01}).
In fact from  Eq.\ (\ref{ageneral}) we have
\begin{eqnarray}
\alpha^{\frac{\gamma_0}{\beta_0}}
&=&
 \left(\frac{L}{\beta_0}\right)^{\gamma_0/\beta_0}
 \left[
  1+\sum_{i=1}^{\infty}\sum_{j=0}^i f_{i,j}{L^*}^i\ln^jL
 \right]^{\gamma_0/\beta_0}
\nonumber\\
&\hspace{-0.7cm}=&  \hspace{-0.5cm}
 \left(\frac{L}{\beta_0}\right)^{\frac{\gamma_0}{\beta_0}}
 \sum_{k=0}^{\infty}
 \frac{\prod_{l=0}^{k-1}(\frac{\gamma_0}{\beta_0}-l)}{k!}
 \left[
  \sum_{i=1}^{\infty}\sum_{j=0}^i f_{i,j}{L^*}^i\ln^lL
 \right]^k
 . \nonumber
 \label{ag1expan}
\end{eqnarray}
Using the square-cup operator, this becomes
\begin{equation}
 \frac{\alpha^{\frac{\gamma_0}{\beta_0}}}
      {\left(\frac{L}{\beta_0}\right)^{\frac{\gamma_0}{\beta_0}}}
=
 \sum_{i=0}^{\infty}\sum_{j=0}^i
 \left(
  \sum_{k=0}^i
 \frac{\prod_{l=0}^{k-1}(\frac{\gamma_0}{\beta_0}-l)}{k!}
  \bigsqcup_{1,0}^k f_{i,j}
 \right)
 {L^*}^i\ln^jL.
\end{equation}
Similarly, we get
\begin{equation} \label{ag2expan}
\sum_{i=0}^{\infty}\tilde{\gamma}_i\alpha^i
=\sum_{i'=0}^{\infty}\sum_{j'=0}^{i'}
 \left(
  \sum_{s=0}^{i'-j'} \grave{\gamma}_s
  \bigsqcup_{0,0}^{s} f_{i'-s,j'}
 \right)
 {L^*}^{i'}\ln^{j'}L,
\end{equation}
where the graved gamma function is given by
\begin{equation}
\grave{\gamma}_s
\equiv\left(\frac{\beta_0}{\beta_1}\right)^s \tilde{\gamma}_s.
\end{equation}
Explicitly, the first several graved beta coefficients are
\begin{eqnarray}
\grave{\gamma_0}
&=& 1,\
\grave{\gamma}_1
=\frac{\gamma_1}{\beta_1}-\frac{\gamma_0}{\beta_0},\
\grave{\gamma}_2
=
 \frac{\beta_0^2}{2\beta_1^2}
 (\acute{\gamma}_0^2+\acute{\gamma}_1),\\
\grave{\gamma}_3
&=&
 \frac{\beta_0^3}{6\beta_1^3}
 \left(
  \acute{\gamma}_0^3+3\acute{\gamma}_0\acute{\gamma}_1+2\acute{\gamma}_2
 \right),\ \cdots
\end{eqnarray}

Substituting Eqs.\ (\ref{ag1expan}) and (\ref{ag2expan})
 into Eq.\ (\ref{mqexpd01}) gives
\begin{equation} \label{mqexpd02}
m_q
=m_{qn} L^{\gamma_0/\beta_0}
 \sum_{i=0}^{\infty}\sum_{j=0}^i
 F_{i,j}
 {L^*}^i\ln^jL,
\end{equation}
where the scale parameters for the quark mass $m_q$ has
been naturally changed to
\begin{equation}
m_{qn}\equiv \hat{m}_q\beta_0^{-\gamma_0/\beta_0}.
\end{equation}
Eq.\ (\ref{mqexpd02}) means that the quark mass is proportional to
$L^{\gamma_0/\beta_0}$ at small $L$ or large $u$, and the proportion
coefficient is nothing but $m_{qn}$.

The expansion coefficients $F_{i,j}$ in Eq.\ (\ref{mqexpd02})
are
\begin{equation}
F_{i,j}
=\sum_{i',j',k,s}
  \grave{\gamma}_s
 \frac{\prod_{l=0}^{k-1}(\gamma_0/\beta_0-l)}{k!}
 \bigsqcup_{1,0}^k f_{i-i',j-j'}
 \bigsqcup_{0,0}^{s} f_{i'-s,j'}
\end{equation}
with the multi-summation being
\begin{equation}
\sum_{i',j',k,s}
=\sum_{i'=0}^i
 \sum_{j'=\mathrm{max}(0,j-i+i')}^{\mathrm{min}(j,i')}
 \sum_{k=0}^{i-i'}
 \sum_{s=0}^{i'-j'}.
\end{equation}
It is also easy to express $F_{i,j}$ as a linear combination
of the graved gamma function, i.e.,
\begin{equation}
\sum_{i',j',k,s}
=\sum_{s=0}^{i-j}
 \sum_{k=0}^{i-s}
 \sum_{i'=s}^{i-k}
 \sum_{j'=\mathrm{max}(0,j-i+i')}^{\mathrm{min}(j,i'-s)}.
\end{equation}
Here are the explicit results to order 4:
\begin{subequations}
\label{Fijconc}
\begin{eqnarray}
F_{0,0}
&=&
 1,\ \
F_{1,0}
=
 \frac{\gamma_1}{\beta_1}-\frac{\gamma_0}{\beta_0},\
F_{1,1}
=\frac{\gamma_0}{\beta_0},\\
F_{2,0}
&=&
 \grave{\gamma}_2
 -\frac{\gamma_0}{\beta_0}
 +\frac{\gamma_0\beta_2}{\beta_1^2},\
\nonumber \\
F_{2,1}
&=&
 \left(\frac{\gamma_1}{\beta_1}-\frac{\gamma_0}{\beta_0}\right)
 \left(1+\frac{\gamma_0}{\beta_0}\right)
 +\frac{\gamma_0}{\beta_0},
 \nonumber\\
F_{2,2}
&=&
 \frac{1}{2}
 \frac{\gamma_0}{\beta_0}
 \left(
  1
  +\frac{\gamma_0}{\beta_0}
 \right), \\
F_{3,0}
&=&
 \grave{\gamma}_3
 +\left(\frac{\gamma_1}{\beta_1}-\frac{\gamma_0}{\beta_0}\right)
 \left(\frac{\gamma_0}{\beta_0}+1\right)
 \left(\frac{\beta_0\beta_2}{\beta_1^2}-1\right)
\nonumber \\
&&
 -\frac{\gamma_0}{2\beta_0}
 +\frac{\gamma_0\beta_0\beta_3}{2\beta_1^3},
\nonumber\\
F_{3,1}
&=&
 \left(2+\frac{\gamma_0}{\beta_0}\right)
 \left(\grave{\gamma}_2+\frac{\gamma_0\beta_2}{\beta_1^2}\right)
\nonumber \\
&&
+\left(1+\frac{\gamma_0}{\beta_0}\right)
 \left(
   \frac{\gamma_1}{\beta_1}-2\frac{\gamma_0}{\beta_0}
 \right),
\nonumber \\
F_{3,2}
&=&
  \left(\frac{\gamma_1}{\beta_1}-\frac{\gamma_0}{\beta_0}\right)
  \left(1+\frac{\gamma_0}{\beta_0}\right)
  \left(1+\frac{\gamma_0}{2\beta_0}\right)
\nonumber\\
&&
  +\frac{\gamma_0}{\beta_0}\left(\frac{3}{2}+\frac{\gamma_0}{\beta_0}\right),
\nonumber\\
F_{3,3}
&=&
 \frac{1}{6} \frac{\gamma_0}{\beta_0}
 \left(1+\frac{\gamma_0}{\beta_0}\right)
 \left(2+\frac{\gamma_0}{\beta_0}\right), \\
&\cdots& \nonumber
\end{eqnarray}
\end{subequations}

These coefficients $F_{i,j}$ satisfy the recursive relation
\begin{equation} \label{Fijrec}
F_{i,j} = \sum_{s=0}^{i-j} F_{i-j,s} {\mathcal R}(s,i,j),
\end{equation}
where ${\mathcal R}$ is a function of three non-negative integers, i.e.,
\begin{equation}
{\mathcal R}(s,i,j)
=\sum_{p=0}^j
 \sum_{t=s}^{j-p}
  \frac{\prod\limits_{l=0}^{p-1}(\frac{\gamma_0}{\beta_0}+l)}{p!}
  \bigsqcup_1^s \frac{1}{t}
  \bigsqcup_0^{i-j}1_{j-p-t}.
\end{equation}
The two square-cup operators,
$\bigsqcup_1^k1_p$ and
$\bigsqcup_0^{i-j}1_{j-t-p}$,
are standard. Their expression can be found in
Eqs.\ (\ref{lna9}) and (\ref{sexmp1}).
%
%
For $i=j$ and $i=j+1$, Eq.\ (\ref{Fijrec}) gives
\begin{eqnarray}
F_{j,j}
&=&
 \frac{1}{j!}\prod_{l=0}^{j-1}\left(\frac{\gamma_0}{\beta_0}+l\right), \\
F_{j+1,j}
&=&
  \left(\frac{\gamma_1}{\beta_1}-\frac{\gamma_0}{\beta_0}\right)
 \prod_{l=1}^j \left(1+\frac{\gamma_0}{l\beta_0}\right)
\nonumber\\
&&
 +\frac{\gamma_0}{\beta_0}
  \sum_{k=1}^j
  \frac{1}{k}\prod_{l=1}^{j-k}
  \left(
   1+\frac{\gamma_0}{l\beta_0}
  \right).
\end{eqnarray}

Truncating Eq.\ (\ref{mqexpd02}) to a finite order
gives a new expression for quark masses, i.e.,
\begin{equation} \label{mqexpd02fin}
m_q
=m_{qn} L^{\frac{\gamma_0}{\beta_0}}
 \sum_{i=0}^{N-1}
 \left(\frac{\beta_1}{\beta_0^2}L\right)^i
 \sum_{j=0}^i
 F_{i,j}
 \ln^jL.
\end{equation}
To leading order, it is simply
\begin{equation}
m_q
=m_{qn} L^{\gamma_0/\beta_0}.
\end{equation}
The next to leading order is
\begin{equation}
m_q
=
m_{qn} L^{\frac{\gamma_0}{\beta_0}}
\left[
1+\frac{\beta_1\gamma_0}{\beta_0^3}L
  \left(
   \ln L
    +\frac{\beta_0\gamma_1}{\beta_1\gamma_0}
    -1
   \right)
\right].
\end{equation}
And the still next to leading reads
\begin{eqnarray}
m_q
&=&m_{qn} L^{\frac{\gamma_0}{\beta_0}}
\Bigg\{
1+L^*
  \left[
   \frac{\gamma_0}{\beta_0}
   \ln L
    +\frac{\gamma_1}{\beta_1}
    -\frac{\gamma_0}{\beta_0}
   \right]
\nonumber\\
&& \hspace{-1cm}
+\frac{{L^*}^2}{2}
\left[
 \left(\frac{\gamma_0}{\beta_0}+\frac{\gamma_0^2}{\beta_0^2}\right)
 \ln^2L
 +2\left(
   \frac{\gamma_1}{\beta_1}
   +\frac{\gamma_0\gamma_1}{\beta_0\beta_1}
   -\frac{\gamma_0^2}{\beta_0^2}
  \right)\ln L
\right.
\nonumber\\
&&\hspace{-1cm}
\left.
 +\frac{\beta_0\gamma_2+\beta_2\gamma_0}{\beta_1^2}
 +\left(\frac{\gamma_1}{\beta_1}-\frac{\gamma_0}{\beta_0}\right)^2
 -\frac{\gamma_1}{\beta_1}-\frac{\gamma_0}{\beta_0}
\right]
\Bigg\}.
\end{eqnarray}

Presently, this process can go to the fourth order
according to the expansion coefficients listed in
Eq.\ (\ref{Fijconc}):
\begin{eqnarray}
m_q
&=&m_{qn} L^{\frac{\gamma_0}{\beta_0}}
\Big[
1+L^*
  \left(
  F_{1,0}+F_{1,1}\ln L
   \right)+
\nonumber\\
&& \hspace{-1cm}
 {L^*}^2
  \left(
   F_{2,0}+F_{2,1}\ln L +F_{2,2}\ln^2L
  \right)+{L^*}^3\left(F_{3,0}
 \right.
\nonumber\\
&& \hspace{-1cm}
 \left.
  +F_{3,1}\ln L+F_{3,2}\ln^2L +F_{3,3}\ln^3L
  \right)
  \Big].
\end{eqnarray}

Now we have three mass formulas, respectively, in
Eqs.\ (\ref{mqexpd02fin}), (\ref{mqexpdfin}), and (\ref{mqexpfinexact}).
To use Eq.\ (\ref{mqexpd02fin}), we need to know $\Lambda$ which
has been listed in Tab.\ \ref{tabLam}, and we take the first row
in the table for each $\Lambda_n$ which has been determined
from Eq.\ (\ref{alfgen3}).
To use Eqs.\ (\ref{mqexpdfin}) and (\ref{mqexpfinexact}), however,
we need to know not only $\Lambda$,  but also the corresponding $\alpha$.
So we naturally use the $\alpha$\ expression in Eq.\ (\ref{alfappr01})
for the mass formula in Eq.\ (\ref{mqexpdfin}), and choose the expression
in Eq.\ (\ref{aexp}) for the mass formula in Eq.\ (\ref{mqexpfinexact}),
in the present numerical calculations.
The next step, we should use an initial condition to determine
the mass scale $m_{qn}$. For $q=t$, $b$, and $c$, the initial
condition is
\begin{equation}
m_q(\bar{m}_q)=\bar{m}_q,
\end{equation}
with $\bar{m}_t=172.76$ GeV, $\bar{m}_b=4.18$ GeV, $\bar{m}_c=1.27$ GeV \cite{PDG2020}.
And for $s$ quark, the initial condition is \cite{PDG2020}
\begin{equation}
m_s(2\ \mbox{GeV})=0.093\ \mbox{GeV}.
\end{equation}
The mass scale $m_{qn}$, calculated from these initial conditions
and the three mass formulas in
Eqs.\ (\ref{mqexpd02fin}), (\ref{mqexpdfin}), and (\ref{mqexpfinexact}),
are listed, respectively, in
Tab.\ \ref{tabmqn}, \ref{tabmqn2}, and \ref{tabmqn3}.
\begin{table}
\centering
\caption{
\label{tabmqn}
 QCD mass scale $m_{qn}$ ($q=s,c,b,t$, and $n=$ 3--6)
from the new expression in Eq.\ (\ref{mqexpd02})
with Eq.\ (\ref{alfgen2}).
        }
\begin{tabular}{ccccccccccc}\hline
 flavor && GeV && $N=1$ && $N=2$ && $N=3$ && $N=4$ \\
  \hline
       &&  $m_{s3}$ && .166190 &&.165242 &&.168878 &&.168611 \\
strange&&  $m_{s4}$ && .180222 &&.177364 &&.181712 &&.181626 \\
       &&  $m_{s5}$ && .205088 &&.198502 &&.203271 &&.203277 \\
       &&  $m_{s6}$ && .249948 &&.237593 &&.243259 &&.243319 \\
     \hline
       &&  $m_{c3}$ && 2.00329 &&1.92027 &&1.97396 &&1.97457 \\
charm  &&  $m_{c4}$ && 2.17243 &&2.06114 &&2.12398 &&2.12698 \\
       &&  $m_{c5}$ && 2.47218 &&2.30678 &&2.37598 &&2.38054 \\
       &&  $m_{c6}$ && 3.01292 &&2.76106 &&2.84339 &&2.84946 \\
     \hline
       &&  $m_{b3}$ && 8.69277 &&8.87212 &&8.98989 &&8.96995 \\
bottom &&  $m_{b4}$ && 9.42671 &&9.52301 &&9.67308 &&9.66232 \\
       &&  $m_{b5}$ && 10.7274 &&10.6579 &&10.8208 &&10.8142 \\
       &&  $m_{b6}$ && 13.0738 &&12.7568 &&12.9494 &&12.9443 \\
    \hline
       &&  $m_{t3}$ && 544.982 &&569.943 &&570.725 &&569.636 \\
 top   &&  $m_{t4}$ && 590.995 &&611.756 &&614.098 &&613.605 \\
       &&  $m_{t5}$ && 672.539 &&684.662 &&686.958 &&686.752 \\
       &&  $m_{t6}$ && 819.645 &&819.492 &&822.098 &&822.029 \\
\hline
\end{tabular}
\end{table}

\begin{table}
\centering
\caption{
\label{tabmqn2}
 QCD mass scale $m_{qn}$ ($q=s,c,b,t$, and $n=$ 3--6)
from the previous expression in Eq.\ (\ref{mqexpdfin})
with Eq.\ (\ref{alfappr01}).
        }
\begin{tabular}{ccccccccccc}\hline
 flavor && GeV && $N=1$ && $N=2$ && $N=3$ && $N=4$ \\
  \hline
       &&  $m_{s3}$ && .195152 &&.170540 &&.170165 &&.168681 \\
strange&&  $m_{s4}$ && .213423 &&.182521 &&.182541 &&.180981 \\
       &&  $m_{s5}$ && .241942 &&.204203 &&.204186 &&.202456 \\
       &&  $m_{s6}$ && .291731 &&.244484 &&.244388 &&.242328 \\
     \hline
       &&  $m_{c3}$ && 2.44972 &&1.98781 &&2.00490 &&1.96575 \\
charm  &&  $m_{c4}$ && 2.67908 &&2.12746 &&2.15072 &&2.10908 \\
       &&  $m_{c5}$ && 3.03708 &&2.38019 &&2.40574 &&2.35934 \\
       &&  $m_{c6}$ && 3.66207 &&2.84970 &&2.87940 &&2.82398 \\
     \hline
       &&  $m_{b3}$ && 9.80497 &&9.08045 &&9.02424 &&8.99662 \\
bottom &&  $m_{b4}$ && 10.7230 &&9.71836 &&9.68058 &&9.65261 \\
       &&  $m_{b5}$ && 12.1558 &&10.8729 &&10.8284 &&10.7980 \\
       &&  $m_{b6}$ && 14.6574 &&13.0176 &&12.9605 &&12.9245 \\
    \hline
       &&  $m_{t3}$ && 576.004 &&574.469 &&572.311 &&572.199 \\
 top   &&  $m_{t4}$ && 629.932 &&614.826 &&613.936 &&613.920 \\
       &&  $m_{t5}$ && 714.109 &&687.865 &&686.733 &&686.767 \\
       &&  $m_{t6}$ && 861.064 &&823.551 &&821.944 &&822.018 \\
\hline
\end{tabular}
\end{table}

\begin{table}
\centering
\caption{
\label{tabmqn3}
 QCD mass scale $m_{qn}$ ($q=s,c,b,t$, and $n=$ 3--6)
from the previous expression in Eq.\ (\ref{mqexpfinexact})
with Eq.\ (\ref{aexp}).
        }
\begin{tabular}{ccccccccccc}\hline
 flavor && GeV && $N=1$ && $N=2$ && $N=3$ && $N=4$ \\
  \hline
       &&  $m_{s3}$ && .195153 &&.173371 &&.170851 &&.169346 \\
strange&&  $m_{s4}$ && .213424 &&.185883 &&.183164 &&.181736 \\
       &&  $m_{s5}$ && .241944 &&.207836 &&.204865 &&.203312 \\
       &&  $m_{s6}$ && .291734 &&.248729 &&.245222 &&.243350 \\
     \hline
       &&  $m_{c3}$ && 2.44973 &&2.07294 &&2.01778 &&1.98684 \\
charm  &&  $m_{c4}$ && 2.67909 &&2.22255 &&2.16321 &&2.13220 \\
       &&  $m_{c5}$ && 3.03709 &&2.48504 &&2.41950 &&2.38534 \\
       &&  $m_{c6}$ && 3.66210 &&2.97398 &&2.89613 &&2.85508 \\
     \hline
       &&  $m_{b3}$ && 9.80494 &&9.10877 &&9.05075 &&9.00491 \\
bottom &&  $m_{b4}$ && 10.7229 &&9.76615 &&9.70307 &&9.66372 \\
       &&  $m_{b5}$ && 12.1558 &&10.9196 &&10.8526 &&10.8110 \\
       &&  $m_{b6}$ && 14.6574 &&13.0680 &&12.9905 &&12.9400 \\
    \hline
       &&  $m_{t3}$ && 575.996 &&573.817 &&573.485 &&572.004 \\
 top   &&  $m_{t4}$ && 629.924 &&615.230 &&614.818 &&613.853 \\
       &&  $m_{t5}$ && 714.099 &&687.890 &&687.659 &&686.732 \\
       &&  $m_{t6}$ && 861.056 &&823.235 &&823.123 &&821.969 \\
\hline
\end{tabular}
\end{table}

\begin{figure}[htb]
\centering
\includegraphics[width=8.2cm]{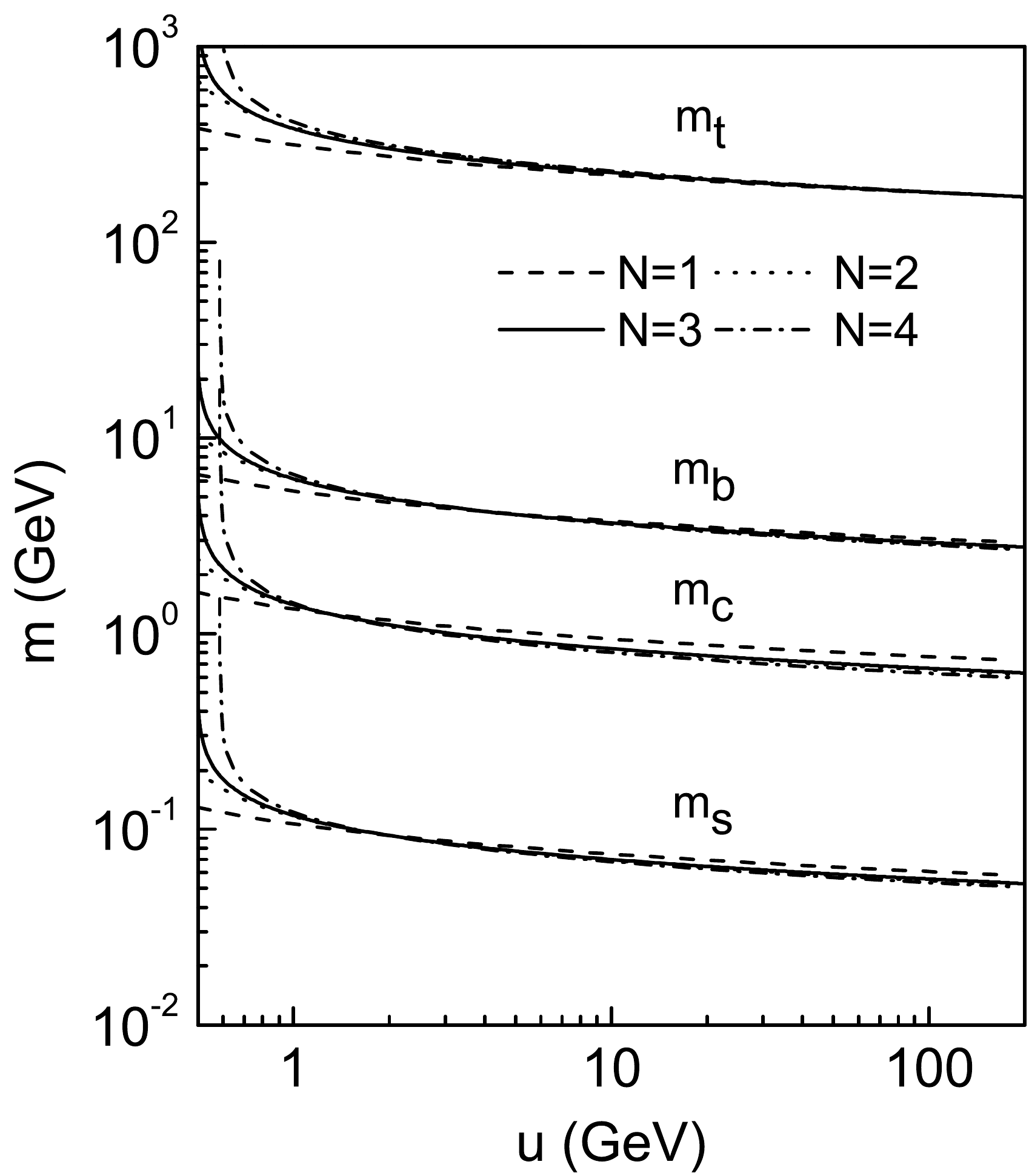}
\caption{
 The running masses of $t$, $b$, $c$, and $s$ quarks as functions
of the RG point $u$,
calculated by the new expression in Eq.\ (\ref{mqexpd02fin})
with $m_{qn}$ in Tab.\ \ref{tabmqn} and $\Lambda$\ in Tab.\ \ref{tabLam},
for the order $N$ from 1 to 4.
         }
 \label{mq}
\end{figure}

Comparing the results in these tables, one can find that the
leading mass scale $m_{qn}$ from the new mass formula
with the new coupling expression is generally closer to
the order-4 value than the two previous formulas.
Figure \ref{mq} shows that the new expression converges indeed
very rapidly. The faster convergence speed can also been seen
from Fig.\ \ref{figam}(b) and \ref{figam}(c), where the leading masses
of $s$, $c$, $b$, and $t$ from the new expression (dashed line)
are compared with that from the previous formulas at
leading order (dotted line) and the order-4 average of the
three formulas (solid line).

\section{Summary}
\label{secsum}

The renormalization group equations for both the QCD coupling and
quark masses have been solved in detailed in a mathematically
strict way. The general relation between the standard expansion
coefficients and the beta and gamma functions are derived.
It is accordingly proved that the fastest convergent expression
can be obtained by simple substitution, which
includes the lost known logarithmic terms and excludes incorrect terms.
At the same time, quark masses can also been expressed as a direct
series of the renormalization point.

It has also been shown that the original beta and gamma coefficients
can be linearly re-combined to give matching-invariant coupling and
quark masses. The concrete matching-invatiant beta and gamma coefficients
are presented to order 4, with a presently incorrect expression for
$\beta_2$ corrected.

Numerical calculations based on the new experimental average
$\alpha_s(M_{\mathrm{Z}})$ has been performed to determined
the QCD scale $\Lambda$\ and quark mass scale $m_{qn}$.
The results show that the new expressions for both the coupling and
quark masses converge indeed much faster.

\section*{Acknowledgments}

The authors would like to thank support from National
Natural Science Foundation of China (Grant No. 11875052, 11575190, and 11135011).

\mbox{}\vspace{0.45cm}
\appendix

\section{square cup operator}
\label{sqcup}

For convenience in the mathematical derivations,
we have introduced the square-cup operator $\bigsqcup$.
The one-dimensional square-cup operator, $\bigsqcup_m^k(a_l)_i$ or
simply $\bigsqcup_m^ka_i$ for short,
is a discrete functional of the three non-negative inters
$k$, $m$, and $i$ and a discrete function $a_l$.
Its value is defined by
\begin{equation}
\left(
 \sum_{i=m}^{\infty} a_i x^i
\right)^k
=\sum_{i=0}^{\infty}
 \left(
 \bigsqcup_m^k a_i
 \right)
 x^i.
\end{equation}
Direct observation gives
\begin{equation}
\bigsqcup_m^0 a_i =\delta_{i,0}, \
\bigsqcup_m^1 a_{i<m}=0, \
\bigsqcup_m^1 a_{i\ge m} =a_i.
\end{equation}
For $k\ge 2$, it is also not difficult to provide a general
explicit expression
\begin{equation}
\bigsqcup_m^k a_i
= \left(
   \prod_{s=1}^{k-1}
   \sum_{p_s=m}^{i-(k-s)m-\varsigma_s^p}
  \right)
   a_{i-\varsigma_k^p} \prod_{r=1}^{k-1} a_{p_r}
\end{equation}
where
\begin{equation}
\varsigma_s^p
 \equiv
\left\{
\begin{array}{ll}
\sum_{t=1}^{s-1}p_t & \mathrm{if}\ s\ge 2 \\
         0          & \mathrm{otherwise}
\end{array}
\right..
\end{equation}
The meaning of $\varsigma_k^p$ is similar to this.
Here are several special simple expressions:
\begin{equation}
\bigsqcup_m^{k>0} a_0=\delta_{m,0}a_0^k, \
\bigsqcup_m^k a_{i<km}=\bigsqcup_m^{k>\frac{i}{m}} a_i=0,\
\end{equation}
\vspace{-0.6cm}
\begin{equation}
\bigsqcup_m^k a_{i<km}
=\bigsqcup_m^{k>\frac{i}{m}} a_{i}
=0,\
\bigsqcup_m^{k>0} a_{km}=a_m^k.
\label{amk}
\end{equation}
If $a_m\neq 0$, the last equality of Eq.\ (\ref{amk})
is also valid for $k=0$.

As an example, let's consider the Taylor coefficients of $[-\ln(1+x)]^k$.
Integrating both sides of $1/(1-x)=\sum_{i=0}^{\infty}x^i$ gives
$-\ln(1-x)=\sum_{i=1}^{\infty}x^i/i$, then
\begin{equation}
\left[-\ln(1-x)\right]^k
=\left[
  \sum_{i=1}^{\infty}\frac{1}{i} x^i
 \right]^k
=\sum_{i=0}^{\infty} \left(\bigsqcup_1^k\frac{1}{i}\right) x^i,
\end{equation}
where $\bigsqcup_1^k\frac{1}{i}$ is a short notation with
the meaning
\begin{equation} \label{lna8}
\bigsqcup_1^k\frac{1}{i}
=\bigsqcup_1^k\left(a_0=0,a_l=\frac{1}{l}\right)_i
=\left\{
 \begin{array}{ll}
  \delta_{i,0} & \mbox{if}\ k=0, \\
  a_i          & \mbox{if}\ k=1, \\
  \bigsqcup_1^{k\ge 2}\frac{1}{i} & \mbox{if}\ k\ge 2,
 \end{array}
 \right.
\end{equation}
and
\begin{eqnarray}
\bigsqcup_1^{k\ge 2}\frac{1}{i}
&=&
 \left(
  \prod_{s=1}^{k-1}
  \sum_{p_s=1}^{i-k+s-\varsigma_s^p}
 \right)
 \frac{1}{i-\varsigma_k^p} \prod_{r=1}^{k-1}\frac{1}{p_r}
\label{lna9} \\
&=&
 \sum_{p_1=1}^{i-1}\frac{1}{p_1(i-p_1)}\ \ \mbox{if}\ k=2,
\label{lna10}\\
&\hspace{-1cm}=& \hspace{-0.7cm}
 \sum_{p_1=1}^{i-2} \sum_{p_2=1}^{i-1-p_1}
 \frac{1}{p_1p_2(i-p_1-p_2)}\ \ \mbox{if}\ k=3,
\label{lna11} \\
 &\cdots& \nonumber
\end{eqnarray}

In the case of $a_l=1$, the multi-summation
can be replaced by compact expressions, e.g.,
\begin{eqnarray}
\bigsqcup_1^{k\ge 2}1_i
&=&
 \bigsqcup_1^{k\ge 2}(a_l=1)_i
 =\frac{1}{(k-1)!}\prod_{t=1}^{k-1}(i-t),
 \label{sexmp1}\\
\bigsqcup_0^{k\ge 1}1_i
&=&
 \bigsqcup_0^{k\ge 1}(a_l=1)_i
 =\frac{(i+k-1)!}{i!(k-1)!}.
\label{sexmp2}
\end{eqnarray}
To have completeness for the square-cup operators
$\bigsqcup_1^k1_i$ and $\bigsqcup_0^k1_i$, we further list
\begin{equation} \label{sexmp3}
\bigsqcup_0^0 1_i=\bigsqcup_1^0 1_i=\delta_{i,0},\
\bigsqcup_1^1 1_0=0, \bigsqcup_1^1 1_{i\ge 1}=1.
\end{equation}

The two-dimensional extension of the square-cup operator,
$\bigsqcup_{m,n}^k(f)_{i,j}$ or $\bigsqcup_{m,n}^kf_{i,j}$ for short,
is defined by
\begin{equation}
\left(
 \sum_{i=m}^{\infty}\sum_{j=n}^i
 f_{i,j} x^i y^j
\right)^k
=\sum_{i=0}^{\infty}\sum_{j=0}^i
 \left(
 \bigsqcup_{m,n}^k f_{i,j}
 \right)
 x^i y^j.
\end{equation}
Similarly we have
\begin{equation}
\bigsqcup_{m,n}^0 f_{i,j}=\delta_{i,0}\delta_{j,0},\
\bigsqcup_{m,n}^1 f_{i<m,j}=0,
\end{equation}
\vspace{-0.6cm}
\begin{equation}
\bigsqcup_{m,n}^1 f_{i,j<n}= 0,\
\bigsqcup_{m,n}^1 f_{i\ge m,j\ge n}=f_{i,j},\
\end{equation}
and for $k\ge 2$, we have
\begin{equation}
\bigsqcup_{m,n}^k f_{i,j}
=\left(
 \prod_{s=1}^{k-1} \sum_{p_s=m}^{i-(k-s)m-\varsigma^p_s}
               \sum_{q_s=\sigma}^{p_s^*}
 \right)
 f_{i-\varsigma^p_k,j-\varsigma^q_k}
 \prod_{r=1}^{k-1} f_{p_r,q_r}
\end{equation}
where
\begin{eqnarray}
p_s^*
&\equiv
&\mathrm{min}
 \left[p_s,j-(k-s)n-\varsigma_s^q\right],
\label{qsup} \\
\sigma
&\equiv&
\mathrm{max}\left(n,p_s+\varsigma_s^p -\varsigma_s^q-i+j\right).
\label{qslow}
\end{eqnarray}
Here are special simple cases:
\begin{equation}
\bigsqcup\limits_{m,n}^k f_{0,0} =\delta_{m,0}\delta_{n,0}f_{0,0}^k,\
\bigsqcup\limits_{m,n}^k f_{km,kn}=f_{m,n}^k,
\end{equation}
\vspace{-0.6cm}
\begin{equation}
\bigsqcup\limits_{m,n}^k f_{i<km,j}
=\bigsqcup\limits_{m,n}^k f_{i,j<kn}=
\bigsqcup_{m,n}^{k>\frac{i}{m}} f_{i,j}
=\bigsqcup_{m,n}^{k>\frac{j}{n}} f_{i,j}=0,\
\end{equation}
\vspace{-0.6cm}
\begin{equation}
\bigsqcup_{m,n}^k f_{i<j,j}=\bigsqcup_{m,n}^k f_{i,j>i}=0,
\end{equation}

From Eqs.\ (\ref{qsup}) and (\ref{qslow}), we can write
\begin{equation}
p_s+\sum_{t=1}^{s-1}(p_t-q_t)-i+j\ge q_s\ge p_s.
\end{equation}
If $j=i$, this means $q_s=p_s$. Therefore, the two-dimensional
square-cup operator reduces to the one-dimensional case, i.e.,
\begin{equation}
\bigsqcup_{m,n}^k f_{i,i}
=\bigsqcup_m^k (a_l=f_{l,l})_i.
\end{equation}

If $f_{i,i}=1\ (i=0,1,2,\cdots)$, which is exactly the case
for the expansion coefficients of the QCD coupling,
one can check the following equality:
\begin{equation}
\bigsqcup_{1,0}^kf_{i,i}
=\bigsqcup_1^k1_i.
\end{equation}
The expression for $\bigsqcup_1^k1_i$ has been listed in
Eqs.\ (\ref{sexmp1}) and (\ref{sexmp3}).
From the above equation, one can further prove
\begin{equation}
\sum_{k=0}^i
\frac{\prod_{l=0}^{k-1}(z-l)}{k!} \bigsqcup_{1,0}^kf_{i,i}
=\frac{1}{i!}\prod_{l=0}^{i-1}(z+l),
\end{equation}
\vspace{-0.6cm}
\begin{equation}
\sum_{i=0}^j\sum_{k=0}^i\frac{\prod_{l=0}^{k-1}(z-l)}{k!}
\bigsqcup_1^k 1_i
=\prod_{l=1}^j\left(1+\frac{z}{l}\right),
\end{equation}
where $z$ is an arbitrary number.

\section{Matching condition for the running coupling}
\label{mat-alf}

In the conventional $\overline{\mbox{MS}}$, the strong coupling
$\alpha(u)$ is not continuous at heavy quark masses.
%
Suppose $\alpha$\ is the coupling in the full
$N_{\mathrm{f}}$-flavor theory while $\alpha^*$
belong to the effective theory with $N_{\mathrm{f}}-1$ flavors.
Then they can be linked by the matching condition:
\begin{equation} \label{mca}
\alpha^*
=\alpha\sum_{i=0}^{\infty}\sum_{j=0}^i
 C_{i,j}\alpha^i\ln^j\frac{m_{\mathrm{h}}^2}{u^2}.
\end{equation}

If the heavy quark threshold $m_{\mathrm{h}}$ is taken to
be running, it should satisfy the RG equation for the quark mass, i.e.,
\begin{equation}
\frac{\mathrm{d}\ln m_{\mathrm{h}}}{\mathrm{d}\ln u^2}
= -\sum_{i=0}^{\infty}\gamma_i\alpha^{i+1},
\end{equation}
where the quark mass anomalous dimension
$\gamma_i$ are given in Eq.\ (\ref{rexp}). 
At the same time, $\alpha^*$ should also satisfies the RG equation
for coupling, i..e,
\begin{eqnarray} \label{rgas}
\frac{\mathrm{d}\alpha^*}{\mathrm{d}\ln u^2}
&=&
  -\sum_{k=0}^{\infty} \beta_k^*{\alpha^*}^{k+2},
\end{eqnarray}
where $\beta_k^*=\beta_k(N_{\mathrm{f}}\rightarrow N_{\mathrm{f}}-1)$.
Substituting Eq.\ (\ref{mca}) into this equation and
taking into account Eq.\ (\ref{RGa}),
the left hand side of Eq.\ (\ref{rgas}) becomes
\begin{eqnarray}
(L)
&=&
  -\sum_{i=0}^{\infty}\sum_{j=0}^i
   \left(
    \sum_{k=0}^{i-j}
     (i-k+1)\beta_k C_{i-k,j}
   \right)
  \alpha^{i+2}\ln^j\frac{m_{\mathrm{h}}^2}{u^2}
    \nonumber\\
&&
 -2 \sum_{i=1}^{\infty}\sum_{j=0}^{i-1}
   \left(
    \sum_{k=0}^{i-j-1}
     (j+1)\gamma_k C_{i-k,j+1}
   \right)
  \alpha^{i+2}\ln^j\frac{m_{\mathrm{h}}^2}{u^2}
    \nonumber\\
&&
  -\sum_{i=0}^{\infty}\sum_{j=0}^i
   (j+1)C_{i+1,j+1}
  \alpha^{i+2}\ln^j\frac{m_{\mathrm{h}}^2}{u^2}.
\end{eqnarray}
and the right hand side of Eq.\ (\ref{rgas}) is
\begin{equation}
(R)
=-\sum_{i=0}^{\infty}\sum_{j=0}^i
 \left(\sum_{k=0}^{i-j}
 \beta_k^*\bigsqcup_{0,0}^{k+2}C_{i-k,j}
 \right)
 \alpha^{i+2}\ln^j\frac{m_{\mathrm{h}}^2}{u^2}.
\end{equation}
Comparing the coefficients of
$\alpha^{i+2}\ln^j(m_{\mathrm{h}}^2/u^2)$ then gives
\begin{eqnarray} \label{Cijeq}
C_{i,j}
&=&
 \frac{1}{j}
\sum_{k=0}^{i-j}
  \left(
   \beta_k^*\bigsqcup_{0,0}^{k+2}
   -(i-k)\beta_k
  \right)
    C_{i-k-1,j-1}
  \nonumber\\
&&
 -2\sum_{k=0}^{i-j-1}
   \gamma_k C_{i-k-1,j},
\end{eqnarray}
which determine all the coefficients $C_{i,j\neq 0}$.
To know $C_{i,0}$, one has to perform perturbative calculations
\cite{Bernreuther1982NPB197,Larin1995NPB438,Chetyrkin1997PRL79,Kniehl2006}.
Here are the results for $C_{i,0}$ to three-loop level
in the modified minimum subtraction scheme
%
%
\begin{subequations}
\begin{eqnarray}
C_{0,0} &=& 1, \ \  C_{1,0} = 0, \ \ C_{2,0} = 11/72, \\
C_{3,0}  \label{C30exp}
&=&
 \frac{575263}{124416}
 -\frac{82043}{27648}\zeta_3
 -\frac{2633}{31104}N_{\mathrm{f}}.
\end{eqnarray}
\end{subequations}
With these values, Eq.\ (\ref{Cijeq}) gives
\begin{equation}
\begin{array}{lccl}
C_{i,j} &=& 0     & \mathrm{if}\ j>i, \\
C_{i,j} &=& 1/6^i & \mathrm{if}\ j=i,
\end{array}
\end{equation}
And for $j<i$, one has
\begin{subequations}
\label{Cijexp}
\begin{eqnarray}
C_{2,1}
&=&  \label{cijstart}
  =11/24
   ,\\
C_{3,2}
&=&
   =\frac{23}{192}-\frac{n}{36}
   ,\
C_{3,1}
 =
  \frac{511}{288}-\frac{67}{576}n
    , \\
C_{4,3}
&=&
  \frac{1597}{10368}
  +\frac{53}{1728} n
  -\frac{1}{324} n^2
    , \\
C_{4,2}
 &=&
  \frac{5317}{6912}
  -\frac{703}{5184} n
  -\frac{77}{20736} n^2
    ,  \\
C_{4,1}
&=&  \label{cijend}
  -\frac{6865}{186624} n^2
  -
  \left(
   \frac{110779}{82944}\zeta_3-\frac{137801}{373248}
  \right) n
 \nonumber\\
&&
  +\frac{2751301}{165888}\zeta_3-\frac{7639841}{746496}
   .
\end{eqnarray}
\end{subequations}
Or explicitly, one has the matching condition
\begin{eqnarray}
\frac{\alpha^*}{\alpha}
&=&1+\alpha t
  +\alpha^2 \left(t^2+\frac{11}{4}t+\frac{11}{72}\right)
   \nonumber\\
&& \hspace{-0.5cm}
  +\alpha^3
   \left[
   t^3
   +\left(\frac{69}{16}-N_{\mathrm{f}}\right)t^2
   +\left(
     \frac{511}{48}
     -\frac{67}{96}N_{\mathrm{f}}
    \right)t
 \right.\nonumber\\
&& \hspace{-0.5cm}
   \left.
 +\frac{575263}{124416}
 -\frac{82043}{27648}\zeta_3
 -\frac{2633}{31104}N_{\mathrm{f}}
   \right]+\cdots
\end{eqnarray}
where $t=(1/6)\ln(m_{\mathrm{h}}^2/u^2)$.

If one would like to use a RG-invariant mass in Eq.\ (\ref{mca}),
then the last term containing $\gamma_k$ in Eq.\ (\ref{Cijeq}) 
should vanish. Accordingly,
the matching condition becomes
\begin{eqnarray}
\frac{\alpha^*}{\alpha}
&=&1+\alpha t
  +\alpha^2 \left(t^2+\frac{19}{4}t+\frac{11}{72}\right)
   \nonumber\\
&& \hspace{-0.5cm}
  +\alpha^3
   \left[
   t^3
   -\frac{131}{16}t^2
   +\left(
     \frac{393}{16}
     -\frac{281}{288}N_{\mathrm{f}}
    \right)t
 \right.\nonumber\\
&& \hspace{-0.5cm}
   \left.
 +\frac{575263}{124416}
 -\frac{82043}{27648}\zeta_3
 -\frac{2633}{31104}N_{\mathrm{f}}
   \right]+\cdots
\end{eqnarray}

In some cases, it would be more convenient to write the matching condition
in Eq.\ (\ref{mca}) in the form \cite{RodrigoPLB424}
\begin{equation}
\alpha
=\alpha^*
 \sum_{i=0}^{\infty}\sum_{j=0}^i
 C_{i,j}^* {\alpha^*}^{i}\ln^j\frac{m_{\mathrm{h}^2}}{u^2}.
\end{equation}
Replacing the $\alpha^*$ here with the right hand side
of Eq.\ (\ref{mca}) gives the relation between
the coefficients $C_{i,j}^*$ and $C_{i,j}$:
\begin{equation}
\sum_{k=0}^i\sum_{l=\max(0,j-k)}^{\min(j,i-k)}
 C_{k,j-l}^*\bigsqcup_{0,0}^{k+1}C_{i-k,l}
=\delta_{i,0}\delta_{j,0},
\end{equation}
Here are the solution to order 4:
\begin{subequations}
\begin{eqnarray}
C_{0,0}^*
 &=&1,\
 C_{1,0}^*=0,\
 C_{i,i}^*=(-C_{1,1})^i,\\
C_{2,0}^*
 &=& -C_{2,0},\
C_{2,1}^*
 =-C_{2,1},\
 \nonumber\\
C_{3,0}^*
 &=& -C_{3,0}, \ C_{3,1}^*=-C_{3,1}+5C_{2,0}C_{1,1},\\
C_{3,2}^*
 &=& -C_{3,2}+5C_{1,1}C_{2,1},\
C_{4,0}^*
 = -C_{4,0}+3C_{2,0}^2,\
  \nonumber\\
C_{4,1}^*
 &=&
  -C_{4,1}+6(C_{3,0}C_{1,1}+C_{2,0}C_{2,1}),\
 \\
C_{4,2}^*
 &=&
  -C_{4,2}+3C_{2,1}^2-21C_{2,0}C_{1,1}^2
   \nonumber\\
&&
    +6(C_{1,1}C_{3,1}+C_{2,0}C_{2,2}),
\\
C_{4,3}^*
 &=&
  -C_{4,3}-21C_{2,1}C_{1,1}^2
   +6(C_{1,1}C_{3,2}+C_{2,1}C_{2,2}).
   \nonumber
 \\
\end{eqnarray}
\end{subequations}

\section{Matching-invariant coupling}
\label{mat-invar-alf}

In the traditional minimum subtraction scheme ($\overline{\mathrm{MS}}$),
the strong coupling $\alpha(u)$ as a function of the
renormalization point $u$\ is not continuous at the
quark masses. In this section, let's derive a matching-invariant coupling
by absorbing loop effects into the $\overline{\mathrm{MS}}$
definition and give the corresponding $\beta$\ function to
four-loop level.

Suppose the new coupling $\alpha'$ is connected to the original
coupling $\alpha$ by
\begin{equation} \label{apdef}
\alpha'
=\sum_{i=0}^{\infty}a_i \alpha^{i+1},
\end{equation}
then,
\begin{eqnarray}
{\alpha'}^*
&=&\sum_{i=0}^{\infty}
 a_i^*{\alpha^*}^{i+1}
 =\sum_{i=0}^{\infty}
 a_i^*
 \left(
  \sum_{j=0}^{\infty}
  C_{j,0}
  \alpha^{j+1}
 \right)^{i+1}
 \nonumber\\
&=&
 \sum_{i=0}^{\infty}
 a_i^*
 \sum_{j=0}^{\infty}
  \left(
   \bigsqcup_0^{i+1}C_{j,0}
  \right)\alpha^{i+j+1}
 \nonumber\\
&=&
 \sum_{i=0}^{\infty}
 \left[
  \sum_{k=0}^i
  a_k^*\bigsqcup_0^{k+1}C_{i-k,0}
 \right]
 \alpha^{i+1}.
\end{eqnarray}
Comparing the coefficients of $\alpha^{i+1}$
in the equality ${\alpha'}^*=\alpha^{\prime}$ leads to
\begin{equation} \label{aisqcup}
a_i
=\sum_{k=0}^i
 a_k^*\bigsqcup_0^{k+1}C_{i-k,0}
\end{equation}

Assume
\begin{equation} \label{aiaij}
a_i
=\sum_{j=0}^i a_{i,j}n^j,
\end{equation}
then $a_k^*=\sum_{j=0}^k a_{k,j}(n-1)^j$.
Substituting into Eq.\ (\ref{aisqcup}) then gives
\begin{equation}
\sum_{k=0}^i
\left[
a_{i,k}n^k
-\left(\bigsqcup_0^{k+1}C_{i-k,0}\right)
 \sum_{j=0}^k a_{k,j}(n-1)^j
\right]=0
\end{equation}
whose solution is
\begin{subequations}
\begin{eqnarray}
a_0 &=& a_{0,0}, \ \
a_1 = a_{1,0},  \ \
a_2 = a_{2,0}+\frac{11}{72}a_{0,0} n, \\
a_3 &=& a_{3,0}
 +\left[
    \left(\frac{7037}{1536}
     -\frac{82043}{27648}\zeta_3
    \right)a_{0,0}
   +\frac{11}{36}a_{1,0}
  \right] n
\nonumber\\
&&
  -\frac{2633}{62208}a_{0,0}N_{\mathrm{f}}^2,\
 \cdots
\end{eqnarray}
\end{subequations}
To definitely fix the new coupling, one needs to choose $a_{i,0}$.
The simplest non-trivial choice would be
\begin{equation}
a_{i,0}=\delta_{i,0}
=\left\{
 \begin{array}{ll}
  1 &  \mathrm{if}\ i=0, \\
  0 &  \mathrm{otherwise}.
 \end{array}
 \right.
\end{equation}
With this convention, one has
\begin{subequations}
\label{aval}
\begin{eqnarray}
&& a_0=1, \ \ a_1=0, \ \ a_2=(11/72) n,
             \label{a012exp}  \\
&& a_3
    =a_{3,1}
     n
    -\frac{2633}{62208} n^2,\
 \label{a3exp}
 \cdots 
\end{eqnarray}
\end{subequations}
where
$
a_{3,1}=7037/1536-82043\zeta_3/27648
\approx 1.014382432.
$
Then the new matching-invariant coupling is
\begin{equation}
\alpha^{\prime}
=
 \alpha
 +\frac{11}{72} n \alpha^3
 +\left(
   a_{3,1}
   -\frac{2633}{62208} n
  \right) n \alpha^4
  +\cdots
\end{equation}

The inverse relation of Eq.\ (\ref{apdef}) is
\begin{equation}
\alpha=\sum_{j=0}^{\infty}
       a_j'{\alpha'}^{j+1}.
\end{equation}
Replacing the $\alpha'$ with the right hand side of Eq.\ (\ref{apdef}),
and then comparing coefficients will give equations to determine
the coefficients $a_j'$:
\begin{equation}
\sum_{k=0}^j
a_k'\bigsqcup_0^{k+1}a_{j-k}=\delta_{j,0},
\ \ \ j=0,1,2,\cdots
\end{equation}
whose solution is
\begin{subequations}
\begin{eqnarray}
a_0'
 &=& \frac{1}{a_0},\ a_1'=-\frac{a_1}{a_0^3},\
a_2'=-\frac{a_2}{a_0^4}+2\frac{a_1^2}{a_0^5},\\
a_3'&=&-a_3/a_0^5+5a_1a_2/a_0^6-5a_1^3/a_0^7,\ \cdots
\end{eqnarray}
\end{subequations}
On application of the $a_i$ values in Eq.\ (\ref{aval}),
these becomes
\begin{equation}
a_0'=1,\ a_1'=0,\ a_2'=-\frac{11}{72}n,\
a_3'=a_{3,1}'n+\frac{2633}{62208}n^2,\ \cdots
\end{equation}
where $a_{3,1}'=-7037/1536+(82043/27648)\zeta_3\approx -1.014382432$.

To transform a thermodynamic potential expressed with
terms like $\alpha^i\ln^j\alpha$, one also needs to know
how to express $\ln\alpha$\ with $\alpha'$\ and $\ln\alpha'$.
Here is the result:
\begin{equation}
\ln\alpha
= \ln\alpha'
 +\sum_{j=1}^{\infty}
 \left(
  \sum_{k=1}^j
  \frac{(-1)^{k-1}}{k}
  \bigsqcup_1^k
  a_j'
 \right)
  {\alpha'}^j.
\end{equation}
%

To three-loop level and with Eq.\ (\ref{aval}), one explicitly has
\begin{equation}
\alpha
=
 \alpha'
 -\frac{11}{72} n {\alpha'}^3
 + n \left(
    a_{3,1}'
   +\frac{2633}{62208} n
  \right)
  {\alpha'}^4
\end{equation}
and
\begin{equation}
\ln\alpha
= \ln\alpha'
 -\frac{11}{72} n {\alpha'}^2
 +n \left(
   a_{3,1}'+\frac{2633}{62208} n
  \right)
  {\alpha'}^3.
\end{equation}

The renormalization equation for $\alpha'$ is
\begin{equation} \label{aprga}
\frac{d\alpha^{\prime}}{d\ln u^2}
=-\sum_{i=0}^{\infty} \beta_i^{\prime}{\alpha^{\prime}}^{i+2}
\end{equation}
The primed beta coefficients $\beta_i^{\prime}$ can be obtained by
substituting Eq.\ (\ref{apdef}) into Eq.\ (\ref{aprga}),
applying Eqs.\ (\ref{aprga}) and (\ref{RGa}), and then comparing
coefficients:
\begin{equation}
\sum_{k=0}^i
\left[
\beta_k^{\prime}\bigsqcup_0^{k+2}a_{i-k}
-(k+1)a_k \beta_{i-k}
\right]=0,
\end{equation}
namely, $\beta_i^{\prime}$ are given
by the recursive relation
\begin{equation}
\beta_i^{\prime} 
=
  \sum_{k=0}^i
   (k+1)a_k \beta_{i-k}
   -\sum_{k=0}^{i-1}
   \beta_k^{\prime}\bigsqcup_0^{k+2}a_{i-k}.
\end{equation}
On application of Eqs.\ (\ref{a012exp}) and (\ref{a3exp}),
one immediately has the explicit expressions:
\begin{subequations} \label{primedbeta}
\begin{eqnarray}
\beta_0^{\prime}
&=& \beta_0=\frac{11}{4}-\frac{n}{6},\
\beta_1^{\prime}
= \beta_1=\frac{51}{8}-\frac{19}{24} n, \\
\beta_2^{\prime}
&=&
 \beta_2+a_2\beta_0
 -a_1(\beta_1+a_1\beta_0)
\nonumber\\
&=&
   \frac{2857}{128}
  -\frac{4549}{1162} n
  +\frac{79}{1152} n^2, \\
\beta_3^{\prime}
&=&
    \beta_3 +2a_3\beta_0
    -2a_1\beta_2+a_1^2\beta_1+4a_1^3\beta_0-6a_1a_2\beta_0
     \nonumber\\
&=&
 \beta_3^{(0)}+\beta_3^{(1)} N_{\mathrm{f}}
    +\beta_3^{(2)} N_{\mathrm{f}}^2
    +\frac{23}{1152}N_{\mathrm{f}}^3,
\end{eqnarray}
\end{subequations}
with
$ 
\beta_3^{(0)}
 = 149753/1536+891\zeta_3/64\approx 114.2303287,\
\beta_3^{(1)}
 = -66733/82944-318179\zeta_3/18432\approx -21.55484044,\
\beta_3^{(2)}
 = -68767/124416+35977\zeta_3/27648\approx 1.011459984.
 \nonumber
$ 

It should be mentioned that a different expression for $\beta_2'$
was previously given in Ref.\ \cite{Marciano1984PRD29}.
The error was caused by the fact that a wrong value for $C_2$
was quoted there \cite{Bernreuther1982NPB197}.


\begin{thebibliography}{99}

\bibitem{Schramm2002}
S. Schramm, W. Greiner, and E. Stein,
{\em Quantum Chromodynamics}, Springer, 2nd edition, 2002.

\bibitem{pgx05EPL72}
G. X. Peng,
 Europhys. Lett. {\bf 72}, 69 (2005);
%
 Nucl. Phys. A {\bf 747}, 75-83 (2005).

\bibitem{Fraga05PRD71}
E.S.\ Fraga and P.\ Romatschke,
 Phys.\ Rev.\ D {\bf 71} (2005) 105014;
E.S.\ Fraga, R.D.\ Pisarski, and J.\ Schaffner-Bielich,
 Phys.\ Rev.\ D {\bf 63} R121702 (2001).

\bibitem{Eidelman2004PLB592.1}
S. Eidelman {\sl et al}. [Particle Data Group Collabration],
 Phy. Lett. B {\bf 592}, 1 (2004).

\bibitem{Chetyrkin1997PRL79}
K.G. Chetyrkin, B.A. Kniehl, and M. Steinhauser,
 Phys. Rev. Lett. {\bf 79}, 2184 (1997).

\bibitem{tHooft1973NPB61.455}
G. 't Hooft,
 Nucl. Phys. B {\bf 61}, 455 (1973);

\bibitem{BardeenPRD18}
W.A. Bardeen, A.J. Buras, D.W. Duke, and T. Muta,
Phys. Rev. D {\bf 18}, 3998 (1978).

\bibitem{pgx06plb634}
G.\ X.\ Peng,
Phys.\ Lett.\ B {\bf 634}, 413 (2006).

\bibitem{Gross1973PRL30.1343}
 D. J. Gross and F. Wilczek,
 Phys. Rev. Lett. {\bf 30}, 1343 (1973);
H. D. Politzer,
 Phys. Rev. Lett. {\bf 30}, 1346 (1973).

\bibitem{Caswell1974PRL33.244}
W.E. Caswell,
 Phys. Rev. Lett. {\bf 33}, 244 (1974);
D.R.T. Jones,
 Nucl. Phys. B {\bf 75}, 531 (1974).
E.S. Egorian, O.V. Tarasov,
Theor. Mat. Fiz. {\bf 41}, 26 (1979).

\bibitem{Tarasov1980PLB93.429}
O.V. Tarasov, A.A. Vladimirov, A.Yu. Zharkov,
 Phys. Lett. B {\bf 93}, 429 (1980);
S.A. Larin and J.A.M. Vermaseren,
Phys. Lett. B {\bf 303}, 334 (1993).

\bibitem{Ritbergen1997PLB400p379}
 T. van Ritbergen, J.A.M. Vermaseren, and S.A. Larin,
 Phys. Lett. B {\bf 400}, 379 (1997).

\bibitem{BurasRMP52}
A. Buras,
 Rev. Mod. Phys. {\bf 52} 199 (1980).

\bibitem{Gardi1998JHEP9807}
E. Gardi, G. Grunberg, and M. Karliner,
JHEP 9807 007 (1998).

\bibitem{Bass2003PRD68}
S. D. Bass, R.J. Crewther, F.M. Steffens, and A.W. Thomas,
Phys. Rev. D {\bf 68} 096005 (2003).

\bibitem{Marciano1984PRD29}
W.J. Marciano,
 Phys. Rev. D {\bf 29} 580 (1984).

\bibitem{PDG2020}
P.A. Zyla et al, [Particle Data Group 2020]
Prog. Theor. Exp. Phys. {\bf 2020}, 083C01 (2020).

\bibitem{Bernreuther1982NPB197}
W. Wetzel,
 Nucl. Phys. B {\bf 196}, 259 (1982);
W. Bernreuther and W. Wetzel,
 Nucl. Phys. B {\bf 197}, 228 (1982);
 Erratum-ibid. B {\bf 513}, 758 (1998);
W. Bernreuther,
 Ann. Phys. (N.Y.) {\bf 151}, 127 (1983);
 Z. Phys. C {\bf 20}, 331 (1983).

\bibitem{Larin1995NPB438}
S.A. Larin, T. van Ritbergen, and J.A.M. Vermaseren,
 Nucl. Phys. B {\bf 438}, 278 (1995).

\bibitem{Kniehl2006}
B. A. Kniehl, A.V. Kotikov, A. I. Onishchenko, and O. L. Veretin,
Phys. Rev. Lett. 97, 042001 (2006).

\bibitem{RodrigoPLB424}
G. Rodrigo, A. Pich, A. Santamria,
 Phys. Lett. B {\bf 424}, 367 (1998).

\bibitem{Chetyrkin1997PLB404}
K.G. Chetyrkin,
 Phys. Lett. B {\bf 404}, 161 (1997).

\bibitem{Vermaseren1997PLB405}
J.A.M. Vermaseren, S.A. Larin, and T. van Ritbergen,
 Phys. Lett. B {\bf 405}, 327 (1997).

\bibitem{Chetyrkin1998npb510}
K.G.\ Chetyrkin, B.A.\ Kniehl, and M.\ Steinhauser,
Nucl.\ Phys.\ B {\bf 510}, 61 (1998).





















%

%
%
%
%

\end{thebibliography}
\end{document}